\documentclass[twocolumn,superscriptaddress,showkeys,amsmath,amssymb,pra,nofootinbib]{revtex4-1}

\usepackage{graphicx}
\usepackage[]{subfigure}
\usepackage{hyperref}
\newcommand{\bitem}{\begin{itemize}}
\newcommand{\eitem}{\end{itemize}}
\newcommand{\benum}{\begin{enumerate}}
\newcommand{\eenum}{\end{enumerate}}
\newcommand{\be}{\begin{equation}}
\newcommand{\ee}{\end{equation}}
\newcommand{\bea}{\begin{eqnarray}}
\newcommand{\eea}{\end{eqnarray}}
\newcommand{\beas}{\begin{eqnarray*}}
\newcommand{\eeas}{\end{eqnarray*}}

\newcommand{\bfr}{{\bf r}}

\newcommand{\equal}{\!=\!}
\newcommand{\lapln}{\nabla^2n}
\newcommand{\gradn}{\nabla n}
\newcommand{\gradnsq}{ \left|\nabla n\right|^2}

\newcommand{\ELF}{\mathrm{ELF}}
\newcommand{\tauTF}{\tau_{TF}}
\newcommand{\tauP}{\tau_{Pauli}}
\newcommand{\tauvW}{\tau_{vW}}

\newcommand{\tauGEA}{\tau_{GEA}}

\newcommand{\tauKS}{\tau_{KS}}
\newcommand{\TKS}{T_{KS}}
\newcommand{\FvW}{F_S^{vW}}

\newcommand{\FGEA}{F_S^{GEA}}

\newcommand{\zfit}{z_{loc}}

\newcommand{\TITLE}{
Visualization and orbital-free parametrization of the large-$Z$
scaling of the kinetic energy density of atoms
}

\begin{document}


\title{ \TITLE }
\author{Antonio C. Cancio}
\affiliation{Department of Physics and Astronomy, Ball State University, Muncie, Indiana 47306}
\email[]{accancio@bsu.edu}
\author{Jeremy J. Redd}
\affiliation{Department of Physics, Utah Valley University, Orem, Utah 84058}
\email[]{accancio@bsu.edu}




\begin{abstract} 
The scaling of neutral atoms to large $Z$, 
combining periodicity with a gradual trend to homogeneity,
is a fundamental probe of 
density functional theory, one
that has driven recent advances in understanding both the kinetic and 
exchange-correlation energies.  
Although research focus is normally upon the scaling 
of integrated energies, insights can also be gained from energy 
densities.
We visualize the scaling of the positive-definite kinetic energy density (KED)
in closed-shell atoms, in comparison to invariant quantities based upon the 
gradient and Laplacian of the density. 
We notice a striking fit of 
the KED within the core of any atom to a gradient expansion using both 
the gradient and the Laplacian, appearing as an asymptotic limit around
which the KED oscillates.
The gradient expansion is qualitatively different from that derived from first 
principles for a slowly-varying electron gas and is correlated 
with a nonzero Pauli contribution to the KED near the nucleus. 
We propose and explore orbital-free meta-GGA models for the kinetic 
energy to describe these features, with some success, 
but the effects of quantum oscillations in the inner shells of atoms makes a 
complete parametrization difficult.  We discuss implications for improved
orbital-free description of molecular properties.
\end{abstract}



\maketitle


\section{Introduction}
The basic insight of density functional theory (DFT)~\cite{HK} is that the ground state 
energy and related quantities are functionals of the particle density alone.
Historically, however, functionals have nearly always been implemented in the 
Kohn-Sham approach which uses auxiliary orbitals derived from the solution 
to an equivalent effective noninteracting problem. 
Orbitals prove very important to describe features in the kinetic energy
such the effect of the quantum oscillations of the shell structure of atoms.
However, the project of developing a true orbital-free DFT, using the density 
only to obtain energies and electronic structure remains a challenge.
This challenge has taken on new impetus with the demand for applications in 
which the use of orbitals is prohibitive~\cite{KCT2013}. 
Such situations include the simulation of mesoscale 
systems~\cite{AkimovPrezhdo} and of 
warm dense matter~\cite{WDM,WDMBasicNeeds} -- matter at high density, at 
temperatures roughly of the fermi temperature, where 
a macroscopic portion of electrons are thermally excited.  
Given robust orbital-free models of exchange
and correlation in the form of generalized gradient approximations (GGA's)
\cite{BeckeGGA,LYP,PBE},
there remains an ongoing need for
developing improved orbital-free models of the Kohn-Sham kinetic energy (KE). 

Much work in this area~\cite{KCST13,KJTH09,APBE,TranWesolowski,LacksGordon94,
Thakkar92}
 has centered on development of GGA's for the KE -- corrections to the 
Thomas Fermi approximation~\cite{Thomas27,Fermi28} constructed from the local density
and its gradient.
These include nonempirical or semi-nonempirical models based on the
satisfaction of exact constraints~\cite{KCST13,APBE}.  A common but not always
accurate~\cite{KarasievConjoint} design principle is that of conjointness with 
exchange~\cite{LeeLeeParr91} -- the development of 
forms that can be adapted to describe both exchange and kinetic energies.
A second area of research is the construction of
nonlocal or two-point functionals, which incorporate 
quantum oscillations such as Friedel oscillations and shell structure 
at the cost of a nonlocal dependence upon density~\cite{WangTeter92,WGC1999,ErratumWGC1999,HuangCarter10,KLXC14}
These have had success for very large solid-state
applications~\cite{HungCarter11,ShinCarter13},
but rely upon material-dependent functionals.

The goal of this paper is to bring together two disparate themes in
density functional theory and bring them to bear upon the problem of 
orbital-free functionals.

The first is as old as density functional theory itself -- the 
large-$Z$ limit of the neutral atom.   As one proceeds down the periodic
table, increasing both nuclear charge $Z$ and electron number 
$N$ to maintain charge neutrality, and allowing both to increase
indefinitely,
one gradually turns off the effects of inhomogeneity on the 
quantum many-body system in a quantifiable way.
The infinite-$Z$ limit for both density and energy is given 
exactly~\cite{LiebSimon,LS73} by the Thomas-Fermi model of the 
atom~\cite{Thomas27,Fermi28,Spruch}, -- a semiclassical 
solution that is essentially a completely orbital-free local density 
approximation.  
The general trend of corrections to this picture as $Z$ is brought down
to realistic values has also long been 
known~\cite{Scott,MarchPlaskett,MarchParr},   
leading to a series expansion in $1/Z^{1/3}$.  These corrections include
gradient corrections to the kinetic energy~\cite{Schwinger80,ES85}  
as well as the introduction of exchange and correlation corrections~\cite{ES82,ElliottBurke,BCGP16} as both vanish relative to the KE 
as $Z\to\infty$.  
At even lower values of $Z$, atoms like third-row transition metals 
provide open challenges to traditional DFT approaches like GGA's and 
meta-GGA's (functionals that use a third variable,
either the Laplacian of the density or the KED in addition to 
the local density and its gradient~\cite{BCGP14,BCGP16}.) 
This scaling thus serves as a natural, disciplined
way to study the gradual introduction of inhomogeneity into density 
functionals.
Essentially, to ascend to large-$Z$ is tantamount to 
descending the Jacob's Ladder of functionals from complicated
orbital-dependent ones to the local-density approximation.

However, it is only 
fairly recently that the implications of this scaling behavior have made 
their way explicitly into density functional development.  Work has been 
done in 
improving the understanding of the connection between the large-$Z$ scaling of 
atomic energies and density functional theory~\cite{PBEsol, ElliottBurke, 
LCPB09, BCGP14, DSFC15, BCGP16}
and along the way, developing new functionals for the kinetic 
energy~\cite{LCFSLapl, APBE, LCPB09, CFDS16}, 
exchange~\cite{PBEsol, APBE, CFDS16},
and most recently, correlation~\cite{BCGP14,BCGP16}.

The other theme in DFT development that we will explore exploits the 
modeling of the Kohn-Sham kinetic energy density -- the 
contribution to the KS KE on a point-by-point basis.  The KED
is an important measure of electronic structure
first of all in a qualitative sense -- as the basis for the electron
localization factor or ELF~\cite{BeckeELF,SavinELF} that identifies 
regions of electron localization such as atomic shells and 
covalent bonds from regions with localized electrons.
It is also the key ingredient in meta-GGA's~\cite{BeckemGGA,MGGA-MS} 
-- where the ELF's ability to diagnose different types of bonds can be used to 
construct functionals that work well for a large variety of systems. 
Recent work on the orbital-free modeling of the KED, and thus implicitly the 
ELF~\cite{CSK16, Finzel15, XC15, XC15comment, XC15reply, LAM14,
PerdewConstantin} demonstrates that 
the gradient and Laplacian of the density taken together can be used to 
construct effective meta-GGA functionals of the KE density.  
This approach has the promise of bringing the insights into electronic 
structure gained from the ELF to the context of OFDFT development. 



This paper is an attempt to combine these two complementary approaches.
Although the KE density of atoms has been the subject of numerous
studies~\cite{Acharya, YPL86, DSFC15, Finzel15}, little
has been done to visualize and analyze their scaling properties
as $Z \to \infty$.
An issue of interest is how different regions of the atom 
scale with $Z$.  There should be a contrast between the interior of the atom 
where the shell structure that characterizes finite atoms tends to the 
smooth Thomas-Fermi limit and 
the near-nuclear core and classically forbidden tunnelling region
far from the nucleus, both of which never converge to the Thomas-Fermi limit.
Particularly, the universal limiting behavior of the KED in these regions
could offer important guidance for functional development as they provide
important boundary conditions that those functionals should try to meet.
A related question is why the gradient expansion works as well as it 
does~\cite{JG}
for these systems despite the significant departures from homogeneity 
in the valence shell and at the nucleus.

In this paper we discuss preliminary results of the visualization of scaling 
behavior of the gradient and Laplacian of the density as a function of $Z$,
and of the Kohn-Sham KED as a function of these quantities.  We show that 
there are at least two types of scaling behavior as $Z$ tends to $\infty$, 
a highly nonanalytic behavior describing the near-nuclear region, 
and the other describable by an empirical gradient expansion in the 
rest of the atom.  Notably, the empirical gradient expansion is different
from that canonically derivable from the slowly-varying electron gas, and 
thus from that used in most GGA and meta-GGA functionals.  This difference
may have significant impact on the ability of these functionals to 
predict binding in molecules.
The rest of this paper is organized as follows: 
Sec.~\ref{sec:theory} 
describes the theoretical background of the paper -- the density functional 
theory of the kinetic energy density, and in particular in the context
of the atomic problem. 
Sec.~\ref{sec:methodology} covers the basic methodology used
for calculations.  
Sec.~\ref{sec:results} 
details the chief results of visualization, and their implications for the
total energy of atoms 
and Sec.~\ref{sec:conclusion} 
presents a discussion of these results and our conclusions.

\section{Theory \label{sec:theory}}

The kinetic energy density in Kohn-Sham theory is given by  
\be
    \tauKS = \frac{1}{2} \sum_i^{occup} f_i \left| \nabla \phi_i \right|^2,
    \label{eq:tauks}
\ee 
where $\phi_i$ are Kohn-Sham orbitals from which the electron density 
is constructed:
\be
      n = \sum_i^{occup} f_i \left| \phi_i \right|^2,
     \label{eq:dens}
\ee
and $f_i$ is the occupation number of each orbital.
Integration over all space gives the kinetic energy
\be
     \TKS[n] = \int \tauKS(\bfr) d^3r.
\ee
A generalization in terms of the spin density and spin-decomposed
KED's may be constructed by restricting the
sums in the equations above to a specific spin species. 
An alternative KED, completely equivalent to Eq.~(\ref{eq:tauks}), is 
\be
    \tauKS' = -\frac{1}{2} \sum_i^{occup} f_i \phi_i^* \nabla^2 \phi_i = 
              \tauKS - \frac{1}{4}\nabla^2 n.
    \label{eq:tauksp}
\ee
Note that the difference is the divergence of a vector function, whose
integral is zero, leaving the integrated KE unchanged.
Eq.~(\ref{eq:tauks}) however is conveniently positive-definite.

A key principle is that $\tauKS$, like 
any other property of an electronic system, is a functional of the 
ground state electron density $n$.  At the same time, this functional
relationship can only be approximated.
A ``semilocal" approximation to $\TKS[n]$ defines $\tauKS$ at 
some position $\bfr$ in terms of the local density, density gradient
and possibly its Laplacian:
\be
         \TKS^{approx}[n] = \int \tau^{approx}[n(\bfr), \gradn(\bfr), \lapln(\bfr)] d^3r
    \label{eq:semilocal}
\ee
~\cite{KCST13,KJTH09,TranWesolowski,LacksGordon94,Thakkar92,PerdewConstantin}.
Another approach, not considered here, involves nonlocal functionals with 
integrals over two 
spatial variables~\cite{JG,WangTeter92,WGC1999,ErratumWGC1999,HuangCarter10}. 

The lowest level of semilocal functional -- the equivalent to the LDA in
XC functionals -- is the Thomas-Fermi model,
\be
   \tauTF = \frac{3}{10} k_F^2 n\!\sim\!n^{5/3},
\ee
with $k_F \!=\! (3 \pi^2 n)^{1/3}$ the fermi wavevector of the homogeneous
electron gas.  
At a next level of approximation is the gradient 
expansion (GEA):~\cite{Kirzhnits57,BrackJenningsChu} 
\be
      \tauGEA = \tauTF + \frac{1}{72}|\nabla n|^2/n + \frac{1}{6}\nabla^2 n 
                    + O(\nabla^4).
      \label{eq:taugea} 
\ee
Terms up to fourth~\cite{Hodges} and sixth order~\cite{Murphy81} in this 
expansion are known.

As is the case with exchange, it is natural 
to recast the derivatives of the density 
into scale-invariant quantities,
here defined as 
\bea
  p &=&  \frac{\gradnsq}{4 k_F^2 n}, \\
  q &=& \frac{\lapln}{4 k_F^2 n}.
   \label{eq:q}
\eea
Then the GEA becomes
\be
       \tauGEA = \left[1 + \frac{5}{27}p + \frac{20}{9}q\right] \tauTF,
      \label{eq:taugeaalso}
\ee
and any generalization of it 
that preserves the proper scaling of $\TKS$ under the uniform scaling of 
the charge density is constructable from an 
enhancement factor $F_S(p,q)$ such that
\be
     \tau_{semilocal} = F_S(p,q) \tauTF.
      \label{eq:taugga}
\ee
Note however even higher order derivatives than $\lapln$ 
may be considered~\cite{dSC15},
but may prove impractical in applications.
The enhancement factor $F_S$ for the kinetic energy plays a role 
equivalent to that for exchange, $F_X$, with $E_X \sim F_X e_X^{LDA}$ 
being the equivalent construction.  The similarities are strong enough
to posit a ``conjointness conjecture"~\cite{LeeLeeParr91}, that the two
enhancement factors $F_S$ and $F_X$ are nearly identical.

For the KED, the most crucial issue for large inhomogeneity $p, q \!\gg\! 1$
is the limit of the one- or two-particle spin-singlet system.
In this case the Kohn-Sham KED reduces to the
the von~Weizs\"acker~\cite{vonWeizsacker} functional:
   \be
       \tauvW = \frac{1}{8} \frac{\left| \nabla n \right|^2}{n},
        \label{eq:tauVW}
   \ee
the exact result for a system of $N$ particles
obeying Bose statistics and having the density $n(\bfr)$.
The KED needed to create $n(\bfr)$ with fermions, the 
energetic cost of Pauli exclusion, 
is given by the difference between the Kohn-Sham and von~Weizs\"acker 
KED's
   \be
    \tau_{Pauli} = \tauKS - \tauvW 
   \ee
from which one can define a Pauli enhancement factor:
\be
   F_{Pauli} = \frac{\tau - \tauvW}{\tauTF},
   \label{eq:fpauli}
\ee
which must hold true for both $\tauKS$ or $\tau^{approx}$.
The Pauli enhancement factor is positive definite:
\be
F_{Pauli} \ge 0
\label{eq:vonweizsacker}
\ee
because of the positive cost of Pauli exclusion~\cite{Herring86}.
Moreover, the response of the fermionic system with respect to 
changes in density must be larger than that of the Bose system:
the Pauli potential 
$\delta T_{Pauli}(\bfr)/\delta n(\bfr) \ge 0$~\cite{LevyHui}. 

Notably, this von~Weizs\"acker lower bound [Eq.~(\ref{eq:vonweizsacker})] is 
not respected by the GEA.
The enhancement factor for $\tauvW$ is $\FvW \!=\! 5p/3$ which gives it 
a coefficient to $p$ that is nine times larger than that of the GEA.  
The resulting Pauli enhancement factor is
\be 
    F^{GEA}_{Pauli} = 1 + \frac{20}{9}q - \frac{40}{27}p
    \label{eq:fgea}
\ee
For $q \equal 0$, (or alternately, dropping the term proportional to $q$
as is done in GGA's) $\tauGEA<\tauvW$ for the relatively modest value of 
$p \equal 27/40$.  

We note here that the gradient expansion correction that is linear in $q$ 
integrates identically to zero.  Thus $q$ will only affect energy expectations 
to fourth order in the gradient expansion. 
The simplest semilocal functionals normally then are constructed as
generalized functions of the remaining variable $p$ -- generalized-gradient
approximations or GGA's.  These
can draw upon a long experience in developing GGA's for exchange and are easy 
to implement. 
The problem is that, even more so than with exchange, functionals at this
level are not flexible enought to be competitive with orbital-dependent models.
Two recent GGA take complementary approaches to address this situation.
The APBE~\cite{APBE}, based on the conjointness conjecture~\cite{LeeLeeParr91} 
takes nearly identical forms for the exchange and 
kinetic energy enhancement factor, and fit both to the large-$Z$ expansion
of atoms to a high degree of accuracy.
This takes advantage of a powerful tool -- 
the scaling of atoms to high-$Z$ is an instance of Lieb-Simon 
scaling~\cite{LiebSimon} in which the effects of inhomogeneity in
a finite system are turned off in a controlled fashion.  
Quite possibly this is an ideal way to construct a 
GGA~\cite{ElliottBurke, APBE, BCGP16}.
The cost of conjointness however, is to break the 
von~Weizs\"acker bound for any finite-$Z$ atom.
The VT84F~\cite{KCST13} imposes both the slowly-varying gas limit
for small $p$ and is limited at large $p$ to $F_{Pauli}(p) > 0$ which 
guarantees the von~Weizs\"acker bound.  Possibly more importantly, 
it guarantees the positive-definite bound on the Pauli potential. 
It has generally however a poor prediction of total KE's~\cite{CSK16}.

A natural way around the problem of conflicting constraints is to put
the extra degree of freedom $q$ back into the functional, that is, to create a 
meta-GGA. 
An instructive attempt is 
the Perdew-Constantin mGGA~\cite{PerdewConstantin}, 
which was developed explicitly to model the kinetic energy density,
as a replacement for the KED in meta-GGA-level XC functionals.  
It starts from  a conventional meta-GGA exact up to fourth order
in the gradient expansion (the GE4-M) to describe the slowly varying limit.  
In order to impose the von~Weizs\"acker bound in the limit of strong
electron localization, it interpolates between this functional and 
the von Weizs\"acker form using a nonanalytic but smooth
function of the difference between the enhancement factors 
$z \!=\! F^{GE4-M} \!-\! \FvW$.
Despite an attractive design philosophy, the mGGA has deficiencies as
a practical tool for OFDFT~\cite{KJTH09,LCFSLapl,CSK16}.
However, it is of value as a an approach for thinking about OFDFT --
building from the basis of the kinetic energy \textit{density} which
is an important tool for visualization and quantitative modeling of 
electronic structure.






Along these lines, perhaps the most physically significant role played by the 
KED in 
a meta-GGA is as a measure of electron 
localization~\cite{BeckemGGA,MGGA-MS,Fam14}.
This is done by taking the ratio of the Pauli contribution to the Kohn-Sham
KED to that of the Thomas-Fermi model,
   \be
         \alpha = \frac{\tauKS-\tauvW}{\tauTF}.
         \label{eq:alpha}
   \ee
In regions where the KE density is determined predominantly
by a single molecular orbital, 
$\tauKS$ approaches $\tauvW$ and $\alpha \!\rightarrow\! 0$.  
This limit describes 
single covalent bonds and lone pairs, and generally situations in which the 
self-interaction errors in the GGA and LDA are most acute.  
The homogeneous electron gas, and presumably systems formed by metallic 
bonds, corresponds to 
$\tauKS~\equal~\tauTF$, $\tauvW\!\sim\!0$ and $\alpha\!\sim\!1$.
Between atomic shells and at low density one finds
$\alpha \!\gg\! 1$, tending to $\infty$ for an exponentially 
decaying density if $\tau_{Pauli}$ vanishes more slowly than $n^{5/3}$.  
This limit can be used to detect weak bonds such as van-der-Waals interactions
and define interstitial regions in semiconductor systems. 
The information on the local environment 
can then be used to customize gradient approximations for specific 
subsystems~\cite{MGGA-MS}.
The electron localization factor or ELF~\cite{BeckeELF,SavinELF} is often 
used in visualization as it converts $\alpha$ into a function with a 
range between zero and one:
   \be
         \ELF = \frac{1}{1 + \alpha^2}.
   \ee
Note that the different contexts developing meta-GGA's and OFDFT's hides an 
important fact: $F_{Pauli} \!= \!\alpha$ for the true Kohn-Sham 
enhancement factor.  Thus
developing an OFDFT is essentially the same problem for both kinetic
and exchange-correlation energies -- that of modeling an orbital-free ELF.

In recent work~\cite{CSK16} we proposed to revise the mGGA following two simple points:
imposing the von~Weizs\"acker lower bound $\tauKS > \tauvW$ 
and relying on the second-order gradient expansion otherwise.  
This satisfies the constraints for the two main limiting cases of the KED -- 
that of delocalized electrons with slowly-varying density
and that of strong electron localization, and otherwise keeps physically 
reasonable behavior for classically forbidden regions with high inhomogeneity. 
We defined a measure of electron localization $z$ as
\be
    z = \FGEA - \FvW - 1 = \frac{20}{9}q - \frac{40}{27}p,
    \label{eq:z}
\ee
which in a sense can be thought of as 
an orbital-free expression for $\alpha$. 

A suitable nonanalytic transition between $\FGEA$ and $\FvW$ may
then be used to impose the von~Weizs\"acker bound, which 
is otherwise broken by the GEA at $z\!\leq\!-1$. 
Adapting a form recently used to construct a $\lapln$-based exchange
function~\cite{CancioExqlaplFull} results in the enhancement factor 
\be
      F^{mGGArev}_S = \FvW + 1 + zI(z),
\ee
where 
\be
      I(z) = \left \{ 1 - \exp{-(1/|z|^\alpha) }\left[1-H(z)\right] \right \}^{1/\alpha}
      \label{eq:mggarev}
\ee
and $H$ is the Heaviside step function.  The interpolation function
$I(z)$ is one for $z>0$ and tends monotonically to $1/|z|$ as $z \rightarrow -\infty$, thus enforcing $F^{mGGArev}_S \rightarrow \FvW$ in this limit.  
Otherwise the functional mimics the GEA, which returns the slowly varying
electron gas for $z\sim 0$, and has the correct scaling behavior for 
$z \rightarrow +\infty$ for a density exponentially decaying to zero.
The differences between this approach and the mGGA are firstly 
the simplification of the functional used in the slowly-varying limit, 
a gradient expansion rather than a meta-GGA.  Secondly the form of 
interpolator between slowly-varying and von~Weizs\"acker limits obeys
a constraint that $\tau$ is greater than  both $\tauGEA$ and $\tauvW$ while 
the mGGA interpolates \textit{in between} the two limits.  
This difference proves to be helpful 
for modeling the KED of covalent bonds~\cite{CSK16}.

The factor $\alpha$ is used to control the rate at which the interpolating
function switches between GEA and vW, with
the leading correction to $\FvW$ being
\be
    \lim_{z\rightarrow -\infty} F^{mGGArev}_S - \FvW \sim \frac{1}{z^\alpha}.
\ee
A factor of $\alpha \!=\! 1$ was considered in the original formulation;
however this changes the value of the cusp in the kinetic energy
density $(d\tauKS(\bfr)/dr)_{r\!=\!0}$ in the vicinity of a nucleus.
%
For hydrogen, this is can be shown to be exactly $-2Z/a_0$, but because
the definition involves taking two derivatives of the particle density, this 
value is not universal. For small atoms it is identical to the cusp condition 
of the von~Weizs\"acker potential, but 
as discussed in the next section, it is altered for larger atoms
by the occupation of $p$-orbitals which have a non-zero 
contribution to the KED at the nucleus.
A safe choice may be $\alpha\! =\! 4$ which does not contribute to the cusp of 
the KED and produces a Pauli potential that is zero at the 
nucleus.  This is presumably the optimal choice for small atoms, like H
where the Pauli KED should be small relative to the von~Weizs\"acker KED,
but possibly not for larger atoms, as the Pauli contribution has to eventually become the dominant piece of the puzzle. 
Finally we note that this approach is not completely new -- 
earlier work of Yang et al.~\cite{YPL86} 
suggested a functional $\tau \!=\! max(\tauvW,\tauGEA)$, essentially the 
$\alpha \to \infty$ limit of the current model.

\subsection{The Kohn-Sham kinetic energy density for atoms}

The radial Kohn-Sham equation for an atom is
\begin{equation}
\label{eq:schrorad}
E_{nl}u_{nl}(r)=\left\{\frac{1}{2}\left[-\frac{d^2}{dr^2}+\frac{l(l+1)}{r^2}\right]-\frac{Z}{r}\right\}u_{nl}(r),
\end{equation}
where $u_{nl}(r)\!=\!rR_{nl}(r)$, $n$ is the principle quantum number, $l$ is the angular momentum quantum number, and $R_{nl}(r)$ is the radial wave function
and $r$ the radial distance from the nucleus.  
The KS density for a closed-shell, spherical atom is given by
\begin{equation}
\label{eq:nofratom}
n(r)= \sum_{l=0}^L\sum_{n=1}^{N} f_{nl} |R_{nl}(r)|^2,
\end{equation}
where $f_{nl}$ is the occupation number for the $n,l$ subshell.
This is strictly correct only for atoms with filled subshells, and we shall
focus on two cases, the noble gases and alkali earths.
%
The kinetic energy density for a spherical atom is
\begin{equation}
\label{eq:KSRKED}
\tauKS= \frac{1}{2}\sum_{l=0}^L\sum_{n=0}^N f_{n,l}\left[
        \left|\frac{dR_{n,l}(r)}{dr}\right|^2 + \frac{l(l+1){R_{n,l}(r)}^2}{r^2}
         \right],
\end{equation}
and the total kinetic energy is
\begin{equation}
\label{eq:KSRKE}
T_{ks}=\int_0^\infty \tau_{ks}(r)d^3r.
\end{equation}

\subsubsection{Scaling to large~$Z$}
An elegant and systematic way of measuring the quality of
approximate density functional theories is test 
their behavior for neutral atoms as the nuclear charge increases.
In the case of hydrogen and helium, 
representing a limit of extreme electron localization,
the KS functional reduces to the von~Weizs\"acker result.  
But as the nuclear charge increases, the core electrons 
of the atom behave more and more like a homogeneous electron gas. 
Thus, for an orbital-free density functional model to predict the 
kinetic energies of any atom, it must be able 
to predict accurately the transition between the homogeneity of extended systems
to the extreme inhomogeneity of small atoms and molecules.  This would then 
make it a good candidate to replace the KS model for a variety of 
systems. 

In the limit of large $Z$, the electronic structure of 
atoms tends exactly~\cite{LiebSimon} to the Thomas-Fermi limit with total 
energy given by $E\!=\!-0.768745 Z^{7/3}$. The density tends nearly 
everywhere to  a universal smooth form, with quantum oscillations due to 
shell structure
decreasing with amplitude as the number of shells increases~\cite{LCPB09}.  
The peak radial 
probability density occurs for $r\!=\! a_{TF}/Z^{1/3}$ with $a$ close to $a_B$; 
with this definition of atomic radius, the atomic radius scales as $Z^{-1/3}$.
The Thomas-Fermi limit describes most accurately the core of the atom where
the density is constructed from many interlacing orbitals and approaches a 
degenerate fermi gas.  It must break down for the innermost shells since
the Thomas-Fermi density unphysically diverges to infinity at $r\!=\!0$; it also
breaks down at large $r$ because the semiclassical approximation used to 
derive the Thomas-Fermi result cannot not describe classically forbidden 
regions.  (In the latter case, the large-$r$ limit of the density decays as
$1/r^6$ rather than exponentially.)

The Thomas-Fermi energy is but the leading 
term in a general asymptotic expansion in $Z$~\cite{LCPB09}. 
For the kinetic energy this 
expansion is known for at least three terms:
\be
   T[Z]= AZ^{7/3} + BZ^{2} + CZ^{5/3} + \cdots.
   \label{eq:largeZ}
\ee
$A\!=\!0.768745$ defines the Thomas-Fermi limit with $T\!=\!-E$ because of 
the virial theorem.  $B\! =\! -0.5$ is the Scott 
correction~\cite{Scott,Schwinger80} which 
corrects the error in the Thomas-Fermi KE caused by the spurious
divergence in the Thomas-Fermi density in the innermost shells 
of an atom.
$C\!=\!0.2699$ defines additional corrections 
derivable from the gradient-expansion correction to the Thomas-Fermi 
picture~\cite{ES85}.

Finally we note that this asymptotic trend is an example of 
Lieb-Simon scaling~\cite{LS73, LiebSimon} where the potential is scaled by 
an arbitrary strength
$\zeta$, distance is scaled by $1/\zeta^{1/3}$, and the number of 
particles in the system is also scaled as $\zeta$ so 
that a charge-neutral system stays charge-neutral.  This
scaling procedure is defined as a generalization of the scaling which occurs 
as one goes down a column of the periodic table.  As it defines the scaling of 
this perhaps most fundamental of all constructs in chemistry, it should be
much more revealing than that of the normal uniform scaling to high
density at fixed particle number.

For the purpose of this paper, we look for three regimes of density, the
large-$r$ asymptotic region $r>Z^{7/6}$, the core of the atom $r\sim Z^{-1/3}$
and the near-nuclear region $r < Z^{-2/3}$.  We should
expect a convergence to the Thomas-Fermi limit, and perhaps the gradient
expansion for intermediate distances, but not for the other two regimes.

\subsubsection{Limits\label{sec:limits}}

A number of facts are known about the KED in the limit of small and 
large $r$, and have recently been characterized 
in some detail~\cite{DSFC15}.  As 
this region defines the leading error in the Thomas-Fermi picture, getting
it right will be important to obtaining good kinetic energies.
Although the density and thus KE density in the core of the atom tends
to a finite value for $r \sim a_0/Z$ or less, the TF charge density
diverges to infinity and the real charge density can never be treated by
this approach.  However, given the vanishingly small role of exchange
and correlation in this limit, one may gain insight by modeling the density
with orbitals taken from the hydrogen atom.

The charge density in this limit is given strictly by the contribution of
$l\!=\!0$ orbitals.  It has the cusp form~\cite{Kato57} for small $r$:
\be
  \lim_{r\rightarrow 0} n(r) \rightarrow n(0)(1 - 2Zr/a_0)
\ee
with $n(0) \sim Z^3/a_0^3$.
This fixes the $r\!=\!0$ value of the von Weizs\"acker KED:
\be
  \lim_{r\rightarrow 0} \tauvW(r) \rightarrow \frac{1}{2}\frac{Z^2}{a_0^2} n(0).
\ee

Taking the atomic KS KED defined above, we decompose into components from
orbitals of specific angular momentum $l$ and sum over all 
shells.  For closed-shell atoms, we obtain
\be
  \tauKS = \sum_l \sum_n \tau_{nl}.
\ee
At the nucleus, $r\!=\!0$, only the two lowest angular momentum components 
contribute: $l\!=\!0$ and $l\!=\!1$.
The $l\!=\!0$ component of the KED is given by
\be
  \tau_0 = \sum_n f_{n0} \left| dR_{n0} /dr\right |^2 = \tauvW.
\ee
The density at the nucleus $n(0)$ is constructed solely from the 
$s$ orbitals and the probability density of each of these 
is of the form $n_{ns}(0)(1 - 2Zr/a_0)$. In other words, each orbital 
separately has the 
limiting cusp condition for the density defined above.   This is enough to show 
that $\tau_0$ is identical to the von~Weizs\"acker model result $\tauvW$.

The $l\!=\!1$ term comes from both non-zero centrifugal energy contribution
to the KED and the square of the derivative of the radial orbital $R_{n1}$.  
It contributes a non-zero Pauli contribution to the KED at the nucleus for 
any atom with at least one occupied $p$ orbital~\cite{Acharya}.
The resulting formula is 
\be
   \tau_1 = \sum_n f_{n1} 3 \left| R_{n1} / r \right |^2 = \tauP.
\ee
As a result, we should expect to find that the $r\!=\!0$ limit of the KED and
more specifically, the Pauli KED, to have a nontrivial dependence on 
the $l\!=\!1$ occupation number and implicitly perhaps upon $Z$.
It is worth noting that it has often been the 
assumption~\cite{PBEsol} that $\tauKS \rightarrow \tauvW$ in 
this limit.  However the true non-zero value of the Pauli KED has
long been known for atoms, and was part of the rationale behind the
construction of functionals using the electron number $N$ about an atom
as an explicit functional variable~\cite{Acharya}. 
The feature has recently been formally characterized and 
generalized to all central-potential problems~\cite{DSFC15},
but it has yet to become part of an effective density functional.

Finally, the large-$r$ limit of $\tauKS$ follows from taking the 
contribution of the HOMO shell to the KED as $r \to \infty$.
For a spherically symmetric atom (a closed shell atom or an open shell atom 
with uniform fractional occupancy), the result is~\cite{DSFC15}
\be
    \lim_{r \rightarrow \infty} \tauKS(r) = \tauvW(r) + \frac{l_H(l_H+1)}{2r^2}n(r)
    \label{eq:tauHOMO}
\ee
where $l_H$ is the angular momentum quantum number of the HOMO shell, and
the particle density $n(r)$ tends to that of the HOMO shell $n_{n_H,l_H}(r)$.
It is notable that neither $\gradnsq$ nor $\lapln$ 
preserves knowledge of the centrifugal force contribution to the KED.  
A radially symmetric density $n(r)$ is constructable without any reference
to the angular components of the Kohn-Sham orbitals so that there is no
way to generate terms that depend upon $l$.  Thus we do not expect a 
good OFDFT model to the Pauli contribution to $\tauKS$ in this limit.

\section{Methodology \label{sec:methodology}}
It is difficult to compare OFDFT models by solving them self-consistently.  
We rather solve the Kohn-Sham equation for a given system and use the 
resulting density for each model.  To this end, 
we use the FHI98PP code~\cite{FHI98PP} to generate Kohn-Sham particle and kinetic energy 
densities.  FHI98PP is an atom code that computes Kohn-Sham orbitals on a 
logarithmic grid of potentially arbitrary accuracy for all particle radii.  
The formula for generating the grid is given by $r_{i+1}=\gamma r_i + r_0$ 
with $\gamma\!=\!0.0247$.  
Because the well-known large-$Z$ expansion is 
nonrelativistic, we do the same for our calculations to be able to make
comparison.
For simplicity, the local density approximation was used for
calculating the exchange-correlation energy.  This does not directly enter into
the calculation of the kinetic energy or kinetic energy density, but might
have some effect on the coefficients of the asymptotic expansion in $Z$.

To calculate the derivatives needed for calculating the KED and the 
Laplacian and gradient of the density on the logarithmic grid, we use a 
Lagrange-interpolation scheme which constructs approximate $n$-th order 
polynomials to be differentiated using $n+1$ grid points. A subgrid of thirteen
 points was found to be optimal, after dropping the first and last six points.  Simpson's method was used for integrals. 

Numerical and analytical tests to determine the accuracy of the differentiation and integration algorithms are described in Ref.~\cite{Redd}. 
The issue of replacing the exact density and LDA density may be assessed
by comparing LDA kinetic energies for noble gases with those obtained using
the optimized effective potential (OEP) method.  These are shown in 
Table~\ref{table:KSerrors}.  Notably, the percent error of the LDA diminishes
rapidly for $Z>10$, as it becomes asymptotically exact for infinite $Z$.

\begin{table}[htb]
\centering
\caption{\label{table:KSerrors} Errors in KS kinetic energies
using the LDA density versus the OEP, from Ref.~\cite{LCPB09}.}
\begin{tabular}{lr|ccr}
\hline\hline
Atom & $Z$ & $T_S$ & $T_{LDA}$ & \% Error \\
\hline\hline
He & 2 & 2.86168 & 2.76739 & 3.295 \\
\hline
Ne & 10 & 128.545 & 127.737 & 0.629 \\
\hline
Ar & 18 & 526.812 & 524.967 & 0.350 \\
\hline
Kr & 36 & 2752.04 & 2747.81 & 0.154 \\
\hline
Xe & 54 & 7232.12 & 7225.09 & 0.097 \\
\hline
Rn & 86 & 21866.7 & 21854.7 & 0.055 \\
\hline\hline
\end{tabular}
\label{table:KSerrors} 
\end{table}

\section{Results \label{sec:results}}

\subsection{Visualizing a parameter space}

\begin{figure*}\centering
      \subfigure[]{
      \label{fig:densAr}
      \includegraphics[width=0.47\linewidth]{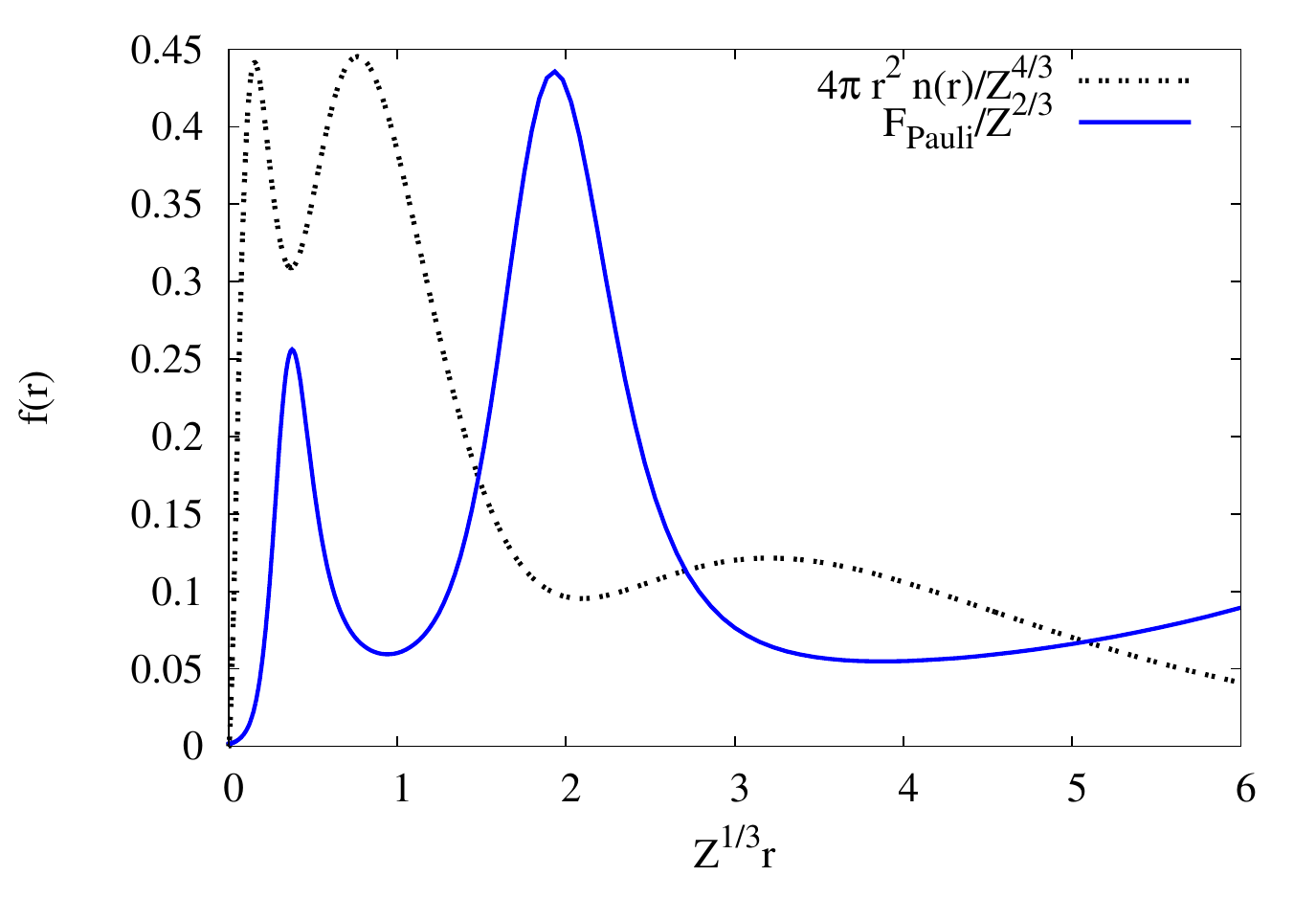}
      }
      \subfigure[]{
      \label{fig:pAr}
      \includegraphics[width=0.47\linewidth]{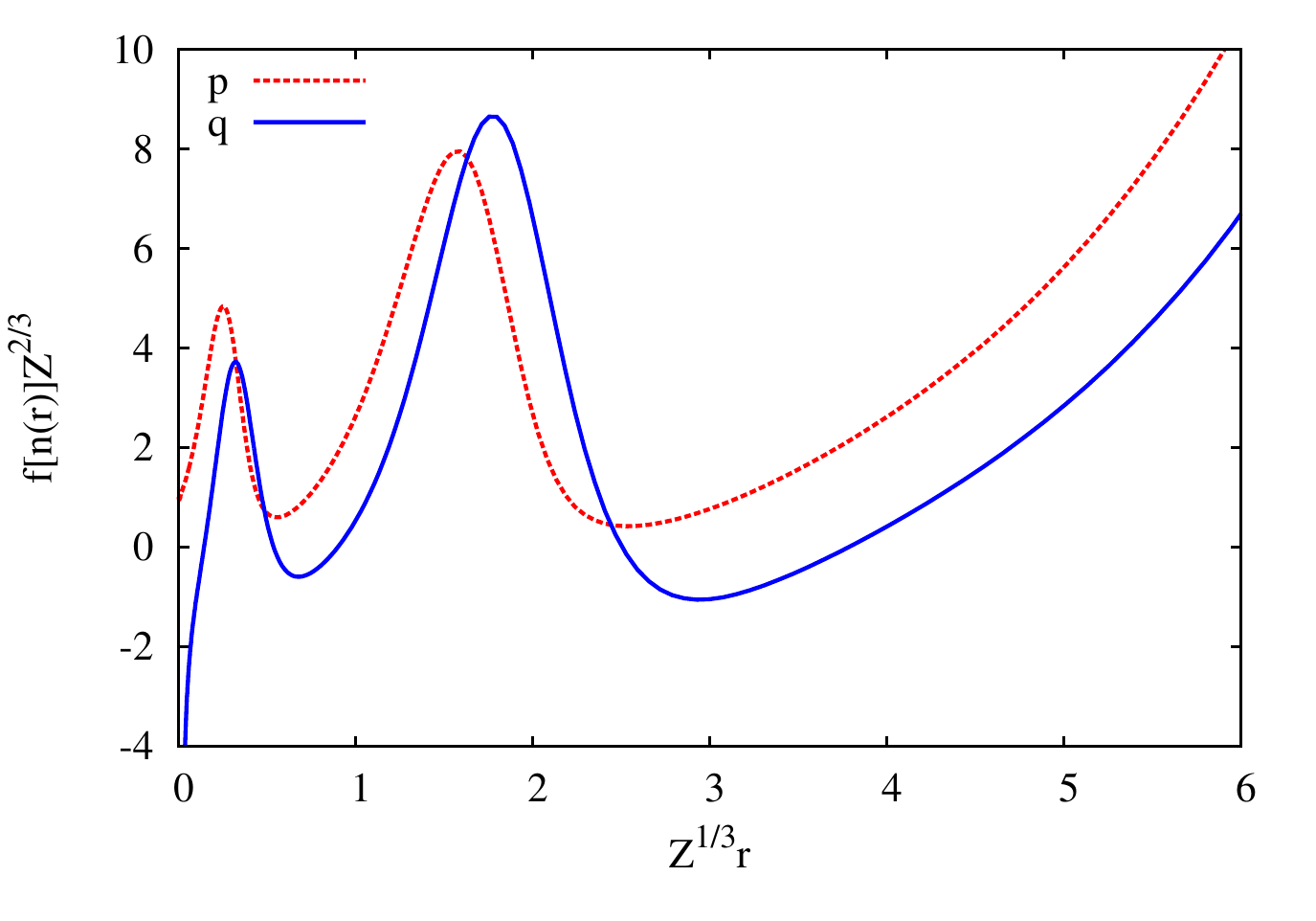}
      }
      \caption{(a) Scaled radial number density $n(r)$ and 
Pauli enhancement factor $F_{Pauli}$ as a function
of scaled radius $Z^{1/3}r$ for Argon. Scaling factors reflect scaling of
atomic peak radius by $Z^{-1/3}$ and particle density by $Z^2$ for 
the Thomas-Fermi atom. (b) $p$ (dotted line) and $q$ (dashed) 
scaled by $Z^{2/3}$ and plotted as function of scaled radius. 
}
\label{fig:Ar}
\end{figure*}

Fig.~\ref{fig:Ar} shows the main players for characterizing the kinetic
energy density of a typical atom, Argon. 
Fig.~\ref{fig:densAr} plots the scaled radial density
versus scaled radius $Z^{1/3}r$.  The peak of the 
Thomas-Fermi density, the $Z\rightarrow\infty$ limit, occurs
at roughly $Z^{1/3}r \!=\! 0.3$~\cite{LCPB09}, in between the $n\!=\!1$ 
and $n\!=\!2$ 
shells; the shells oscillate above the TF peak value of $\sim 0.38$.
Fig.~\ref{fig:pAr} shows suitably scaled values of $p$ and $q$ versus
scaled radius.  As noticed by Bader in the development
of the QTAIM~\cite{BaderReview,BaderEssen84}, the Laplacian of the 
density, proportional to $q$, is negative (or more reliably, at a local
minimum) at the centre of each shell, and is a local maximum in between
shells.  It tends to $-\infty$ at the nucleus
because of the cusp in the electron density and to $+\infty$ far from the atom.
The gradient variable $p$ is finite at $r\!=\!0$ but otherwise shows a
similar behavior as $q$, with $q$ lagging slightly behind it
in a way reminiscent of sine and cosine functions.

\begin{figure}[!htbp]\centering
      \includegraphics[width=0.9\linewidth,height=0.41\textheight,keepaspectratio]{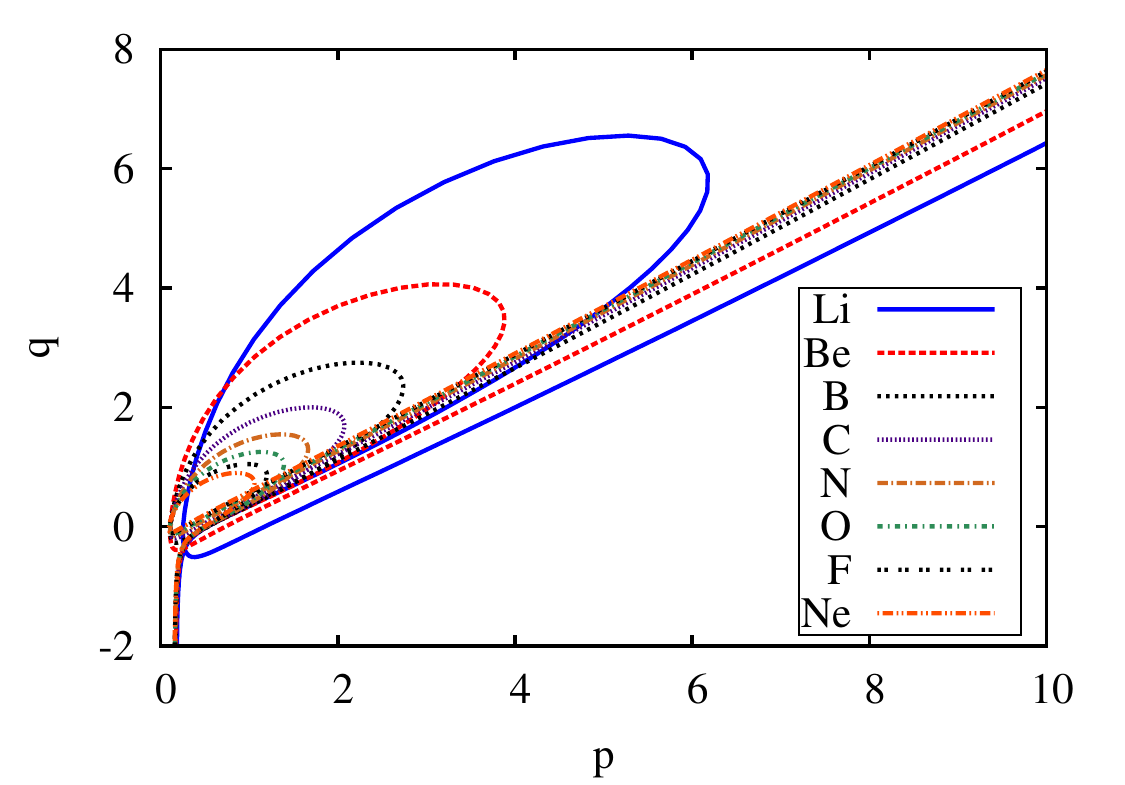}
      \caption{\label{fig:row2}Parametric plot of $p(\bfr)$ vs $q(\bfr)$ for row two of the periodic table.}
\end{figure}

\begin{figure}[!htbp]\centering
      \includegraphics[width=0.9\linewidth,height=0.41\textheight,keepaspectratio]{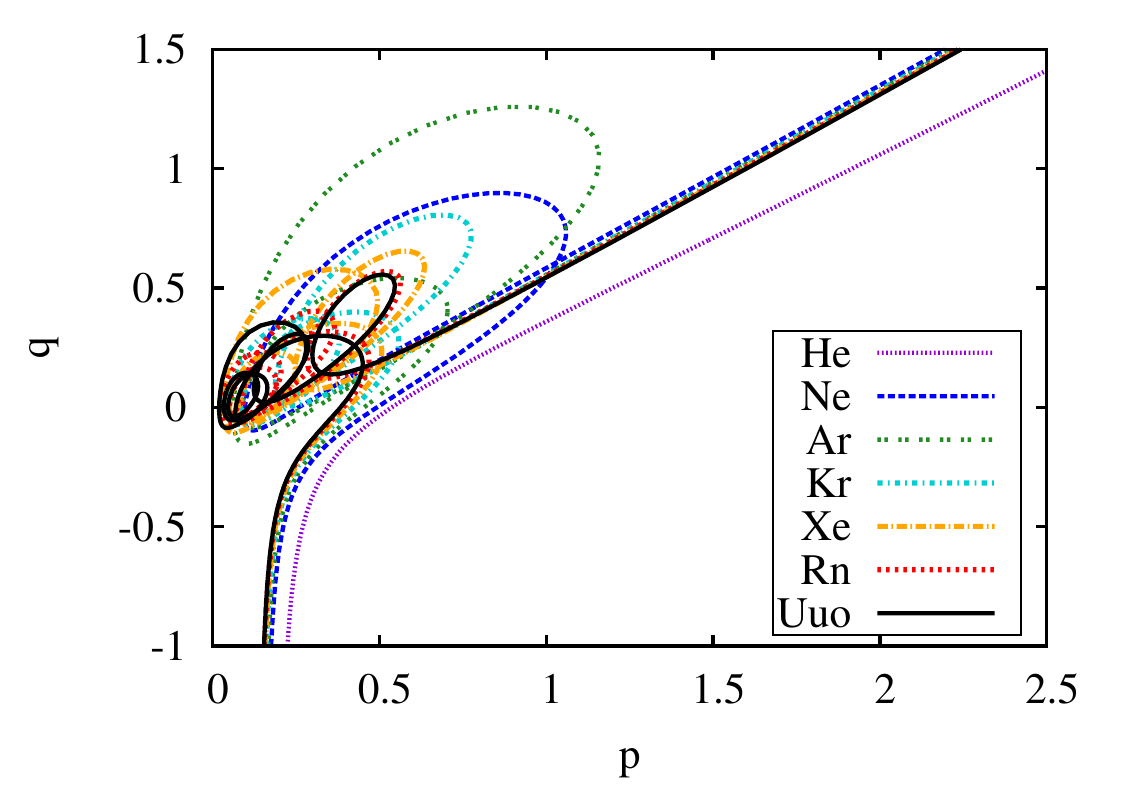}
      \caption{\label{fig:3regions}Parametric plot of $p(\bfr)$ vs $q(\bfr)$ for all atoms in column VIII of the periodic table.}
\end{figure}

We may gain more insight by plotting $q(r)$ versus $p(r)$, an analog to 
the phase-space plot 
$d\theta(t)/dt$ versus $\theta(t)$ encountered in the study 
of oscillator dynamics.  The results for the first row of the periodic 
table, from Li through Ne, are shown in Fig.~\ref{fig:row2} and for the 
noble gases in Fig.~\ref{fig:3regions}.
Comparing to Fig.~\ref{fig:pAr}, we can identify the three pertinent 
regions of the atom as three distinct features in ``phase-space."
The classically-forbidden asymptotic region far from the nucleus shows up
as a linear tail that extends to positive infinity in both $p$ and $q$.  
The region
near the nucleus characterized by the cusp in the electron density is the other
end of each phase-space ``trajectory", where $q \rightarrow -\infty$ and
$p$ is finite and varies little with $Z$.
A system with only one shell, such as He in Fig.~\ref{fig:3regions}, transitions
from the one region to the other seamlessly.  Otherwise there is
exactly one loop in $p$ and $q$ for every shell transition.  The $n\!=\!2$ to
$n\!=\!1$ or L to K shell transition is observable in Fig.~\ref{fig:row2};
close observation of Fig.~\ref{fig:3regions} reveals one loop for Ne, two for
Ar, three for Kr and so on.
The largest $p$ and $q$ values occur in the transition between shells, and
the smallest at valence shell peaks. Thus 
in the midregion between the two extremes of cusp
and asymptote, there is a tendency towards weak 
relative gradient corrections $p,q \ll 1$ -- 
that is, towards the slowly-varying electron gas. 

The trend to infinite $Z$ in this picture is also revealing.  
The behavior of $p$ and $q$ in the cusp and asymptotic regions
is essentially unchanging -- there is only a 
modest shift from the He atom case to the largest $Z$ atom.  This may reflect 
the fact that neither of these two regimes can be adequately described in 
Thomas-Fermi theory: the charge density is singular at the nucleus and 
decays as $1/r^6$ as $r \rightarrow \infty$.  One sees in some sense
a renormalization of the trend described by the Helium atom -- that is 
of the atomic features of the system furthest from the TF limit.
It is in the core shells of the atom, which should eventually trend to the 
TF limit that a dependence upon $Z$ is most clearly seen.
The trend down the first row, shown in Fig.~\ref{fig:row2}, is 
of the shell structure loop transitioning from an exceptionally 
large range of $p$ and $q$ for the smallest-$Z$ atom, slowly towards 
the $p\!=\!q\!=\!0$ limit.  By Neon, the majority of the atom is within the 
range $p, q \!<\! 1$.  

As further shells are added onto the system 
(Fig.~\ref{fig:3regions}), 
the space for any particular 
transition -- L to K, M to L, N to M -- consistently shrinks. 
Interestingly, the second innermost loop caused by the transition from
the M to L shells rapidly shrinks to the perturbative regime $p,|q|\ll1$ -- 
one rapidly reaches the slowly-varying limit for inner shells as
predicted by TF theory.  However, the last transition, between K and L causes
a large swing-out to higher $p$ just before the trajectory transitions
to the nuclear cusp.  This may be indicative of the argument behind the 
Scott correction to the KE (the second term in Eq.~\ref{eq:largeZ}) -- 
that it involves not only the 1s shell, but contributions
from the other innermost shells as well~\cite{Schwinger80}. 
Focusing on the HOMO shell, the trend is less predictable but follows 
very gradually to the slowly-varying limit.  

\subsection{Parametric visualizaiton of the kinetic energy density}
\begin{figure*}[!htbp]\centering
      \subfigure[]{
      \label{fig:3d_30}
      \includegraphics[width=0.47\linewidth]{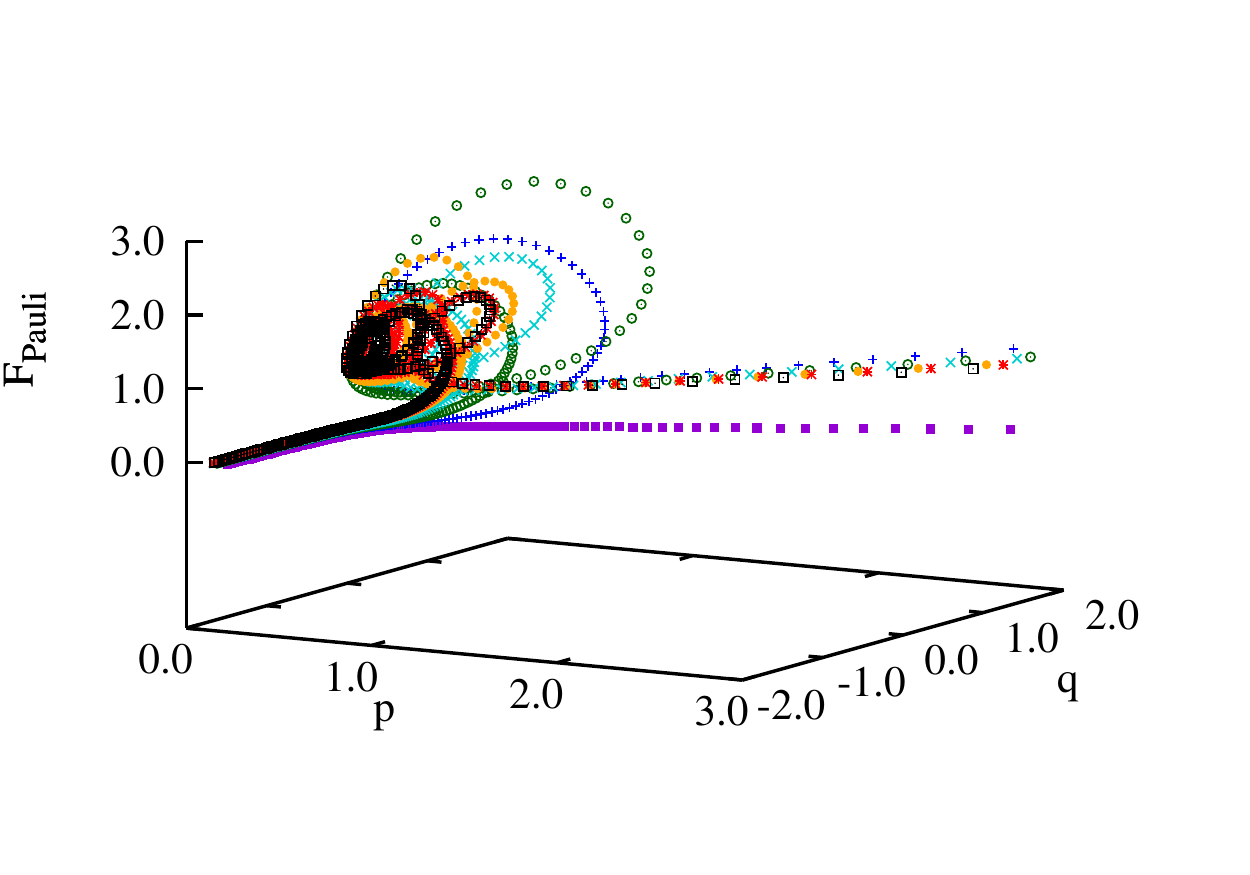}
      }
      \subfigure[]{
      \label{fig:3d_120-eps-converted-to.pdf}
      \includegraphics[width=0.47\linewidth]{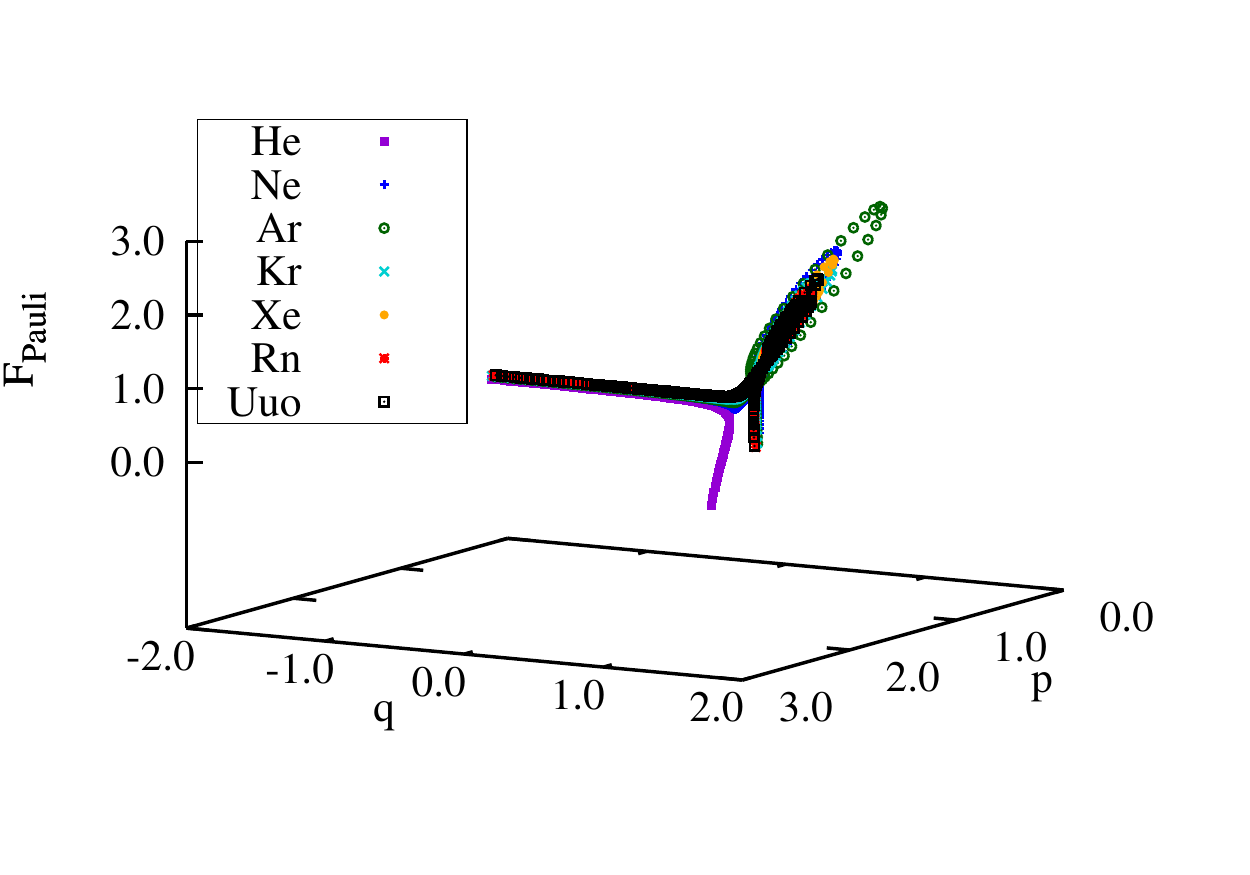}
      }
      \caption{(a) $F_{KS}^{Pauli}(\bfr)$ versus $p(\bfr)$ and $q(\bfr)$ 
        for noble gas atoms.  Perspective is rotated 30 degrees about the 
        $z$-axis with respect to the $p$-axis.  (b) Same, for a 120 degree 
        rotation.}
      \label{fig:3d}
\end{figure*}

Up to now only the visualization of the space defined by 
$p(\bfr)$ and $q(\bfr)$ has been discussed. 
We now include the Pauli enhancement factor of the Kohn-Sham
KED, given by 
Eq.~(\ref{eq:fpauli}) in the third dimension.  
The result for the noble gases is shown as a scatter-plot over 
the numerical logarithmic grid in Fig.~\ref{fig:3d_30}. 
This results in a three-dimensional parametric plot similar to the 
two-dimensional plot in Fig.~\ref{fig:3regions}. 
The view is rotated $30^{\circ}$ about the $z$ axis in Fig.~\ref{fig:3d_30}
and $120^{\circ}$ in (b). 

Note that He, shown as violet circles, has zero Pauli KED and thus lies
entirely in the $F_{Pauli}\!=\!0$ plane.  All parametric curves start with
a nearly universal behavior with $F_{Pauli} \sim 0$ near the nuclear
cusp, shown as the tail for $p \sim 0$ and $q<0$.  The noble gases 
show approximately the same behavior for very large $r$, forming a second
nearly universal curve.  This however shows distinct signs of fanning out
and is significantly different from He, or for other atoms, like Be, with
no $p$ frontier orbitals.

Most remarkably,
it can be seen especially from (b) that the frontier and core regions 
of every atom are nearly coplanar.   There is a perspective, not too
far from that shown in (b) which looks
at that plane edge on, in which the whole parametrized
enhancement factor over all noble gases reduces to a simple hockey-stick
form.
This has several implications.  For the observable range of values
of $p$ and $q$, $F_{Pauli}$ for the noble atoms reduces to nearly a 
single-valued function of the two variables $p$ and $q$.  While either
separately might lack sufficient information to characterize this set
of systems, the combination does, and thus an unambiguous orbital-free
functional may be constructed.    
But more than this: over much of its range, $F_{Pauli}$ reduces to a 
simple linear function of the two.  In terms of density functional theory,
the Pauli enhancement factor is in large part that of 
a second-order gradient expansion.  Finally, the region
of the parameter space where the $F_{Pauli}$ data does \textit{not} fall into
 a plane is that of the cusp in the density near the nucleus, where
a different universal behavior holds.  The net result is that both regions 
can be described by a single parameter -- a linear combination of $p$ and 
$q$.  The determination of this parameter and its use in modifying 
density functionals is described in the next sections.

\subsection{Gradient Expansion Fits}


We now assume the projection of $F_{Pauli}$ 
onto a function 
defining a plane in 
${p,q}$ space.
This describes a fit to a GEA: 
\be
 F_{Pauli}^{GEAloc} = 1 + \zfit
  \label{eq:geafit}
\ee
with 
\be
    \zfit = (a \cos{\theta}) p + (a \sin{\theta}) q
   \label{eq:zfit}
\ee
being an empirical version of the $z$ variable introduced in Eq.~(\ref{eq:z}).
This defines a GE valid locally for the KE density rather than the normal
GE, derived for the KE.
Then $a$ and $\theta$ can determined by a least-squares-fit over a suitable
range in $p$ and $q$.  

Ideally, given that the GEA 
should be most applicable in the limit $Z\rightarrow \infty$, we should take 
an extrapolation to the largest-$Z$ atom numerically feasible.  Such
calculations of 1000's of electrons are chemically unrealizable but 
mathematically important for accurately determining limiting 
cases~\cite{BCGP16,APBE}.  Secondly, we should limit the range 
of the fit to values of $p, |q| \ll 1$, the range of validity for the
gradient expansion.

\begin{figure*}[!htbp]\centering
      \subfigure[]{
      \label{fig:fittrend_a}
      \includegraphics[width=0.47\linewidth,height=0.22\textheight,keepaspectratio]{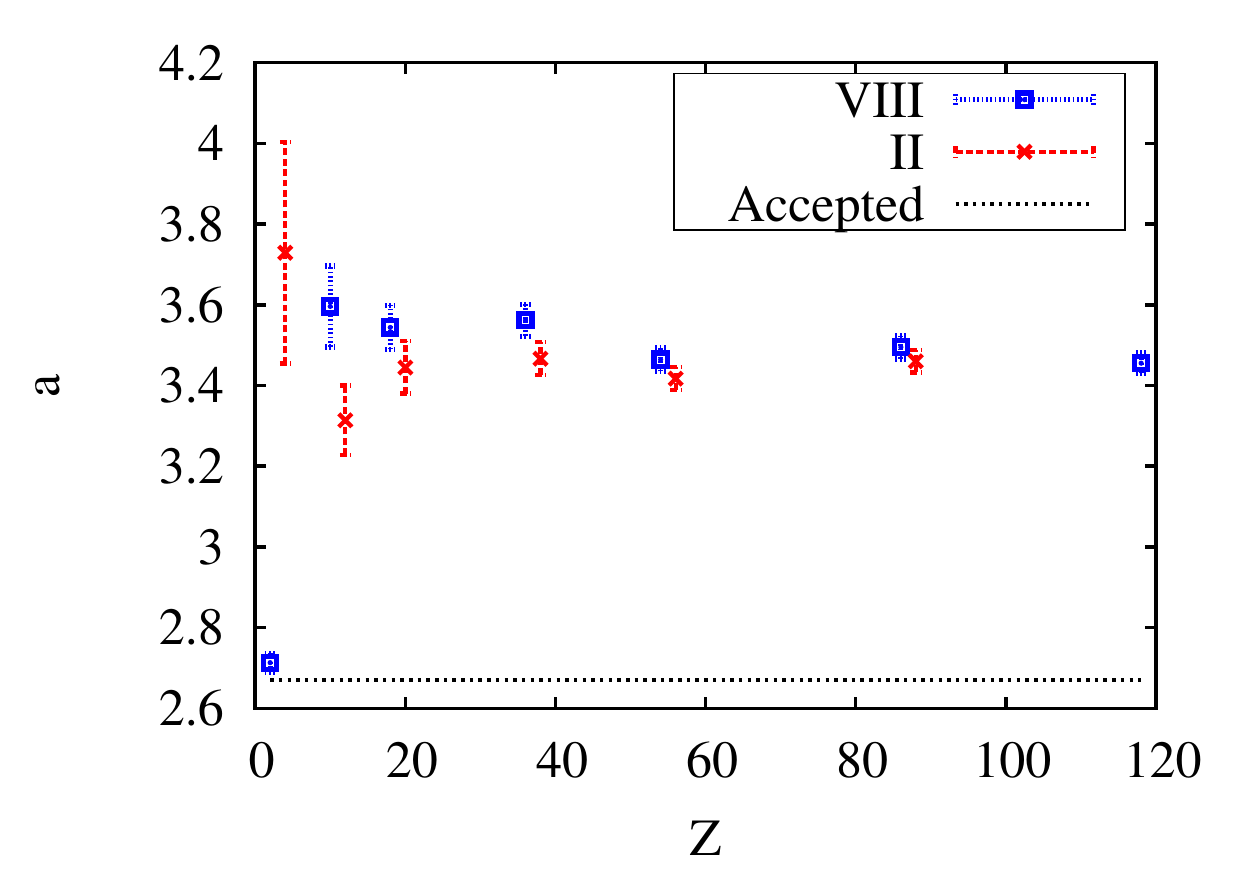}
      }
      \subfigure[]{
      \label{fig:fittrend_theta}
      \includegraphics[width=0.47\linewidth,height=0.22\textheight,keepaspectratio]{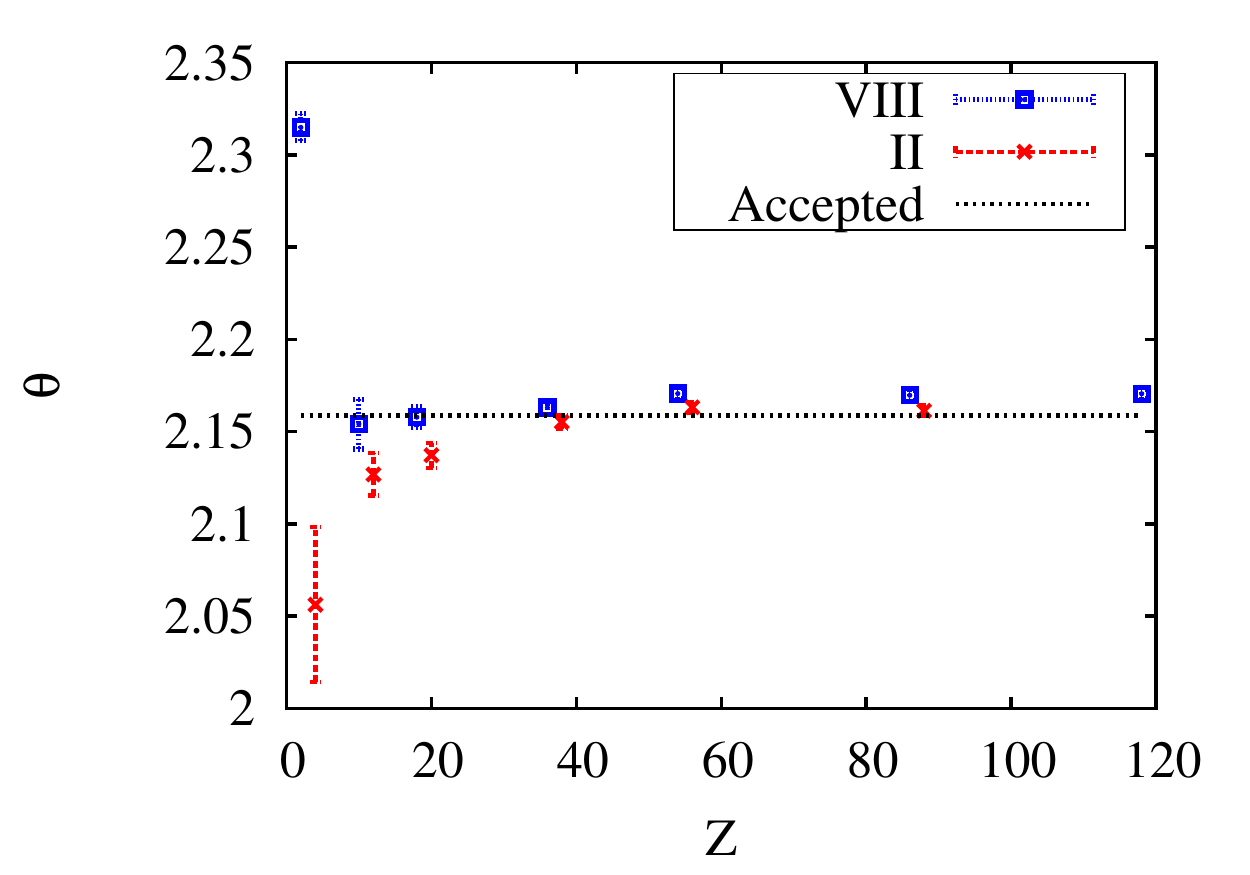}
      }
      
      \caption{ \label{fig:fittrend}
(a) Fit parameter $a$ and (b) fit parameter $\theta$ versus 
$Z$ as determined by fitting Eqs.~(\ref{eq:geafit}) 
and~(\ref{eq:zfit}) to $F^{KS}_{Pauli}(p,q)$ for 
individual atoms.  These are compared to values of $a$ and $\theta$ 
from conventional gradient expansion (dotted line.) } 
\end{figure*}

A preliminary calculation shows that this may not be too important for 
our purposes.  
We perform a least squares fit of
$F_{Pauli}$ to Eqs.~(\ref{eq:geafit}) and~(\ref{eq:zfit}) 
for a given atom over all numerical
grid points $r_i$ for which
$p(r_i) < 0.6$ and $-0.125 < q(r_i) < 0.6$.
The results are shown for the alkali earths and noble gases in 
Fig.~\ref{fig:fittrend_a} for $a$ and 
\ref{fig:fittrend_theta} for $\theta$.  The results converge very nearly to 
a constant for both columns after about $Z\!=\!50$.  Taking the 
last five atoms shown and averaging we get 
$a \!=\! 3.459(13) $ and 
$\theta \!=\! 2.1652(13)$.  
Taking the data for Uuo ($Z\!=\!118$) only, and restricting the fit further
to $p,q\!<\!0.5$, 
we get $a \!=\! 3.486(26)$ and 
$\theta \!=\! 2.1615(28)$,
a near match.

One point of interest here is that the values found empirically do not
match those of the canonical~\cite{Kirzhnits57} gradient expansion.  
The corresponding 
values of $a$ and $\theta$ obtainable from Eq.~(\ref{eq:fgea}),
$a \!=\! 2.671$ and $\theta \!=\! 2.159$, 
are shown as straight lines in Fig.~\ref{fig:fittrend}.
Apparently, $\theta$, measuring the relative mixture of $p$ and $q$ 
to the gradient expansion correction to the KED is unchanged to 
within statistical error.
However the magnitude of the GE correction $a$ converges quickly with $Z$
 to a value 30\% larger than the predicted correction.

That is to say, the actual gradient expansion of the KE \textit{density},
within the core region of the atom where this expansion is locally valid,
is not the gradient expansion of the \textit{integrated} KE.

The implications of this difference are quite dramatic.  Convert these
parameters back to the expression Eq.~(\ref{eq:taugga}) for the KED and
then to an expression for the total KE.  We then get the following expressions
for the result produced by the empirical local GEA for Uuo and the
canonical GEA:
\bea
     T^{GEA} &=&    \int d^3r (1 + 0.185 p + 2.222 q) \tauTF \\
     T^{GEAloc} &=& \int d^3r (1 - 0.275 p + 2.895 q) \tauTF.
     \label{eq:TGEAloc}
\eea
Given that for a pure GEA functional, the GE term linear in $q$ integrates
to zero, the net GE contribution to the kinetic energy from the local
GEA fit is the opposite sign from that of the canonical GE.
As we shall see further on, it is actually the \textit{wrong} sign --
giving a GE expression for the energy that is worse than that for 
the Thomas-Fermi model.
 
It is also interesting that this is not the first evidence of such a qualitative
discrepancy between the gradient expansions of the KE and KED.
The recent analytic gradient expansion of the KED of 
the Airy gas~\cite{LAM14}, a system
that asymptotically approaches an electron gas with a constant density gradient,
also produces a negative coefficent for $p$.  
In this case, the kernel for the KE integral is $F_S = 1 - 0.185 p + 3.333 q$,
which shows a similar change from the standard gradient expansion as that
of the atom.  However, quantitatively, these numbers are far outside the
error bars of our statistical fits for the atom -- the asymptotic limit of 
the KED of the neutral atom clearly tends to a different gradient expansion 
than that of the Airy gas.
Nevertheless, it is reasonable to say that the gradient
expansion about the local density approximation limit of a sloped system,
either atom or Airy gas, is fundamentally different from that about the 
homogeneous electron gas.

\subsection{Single-variable projection of the KED}
\begin{figure}[!htbp]\centering
      \includegraphics[width=0.9\linewidth,height=0.40\textheight,keepaspectratio]{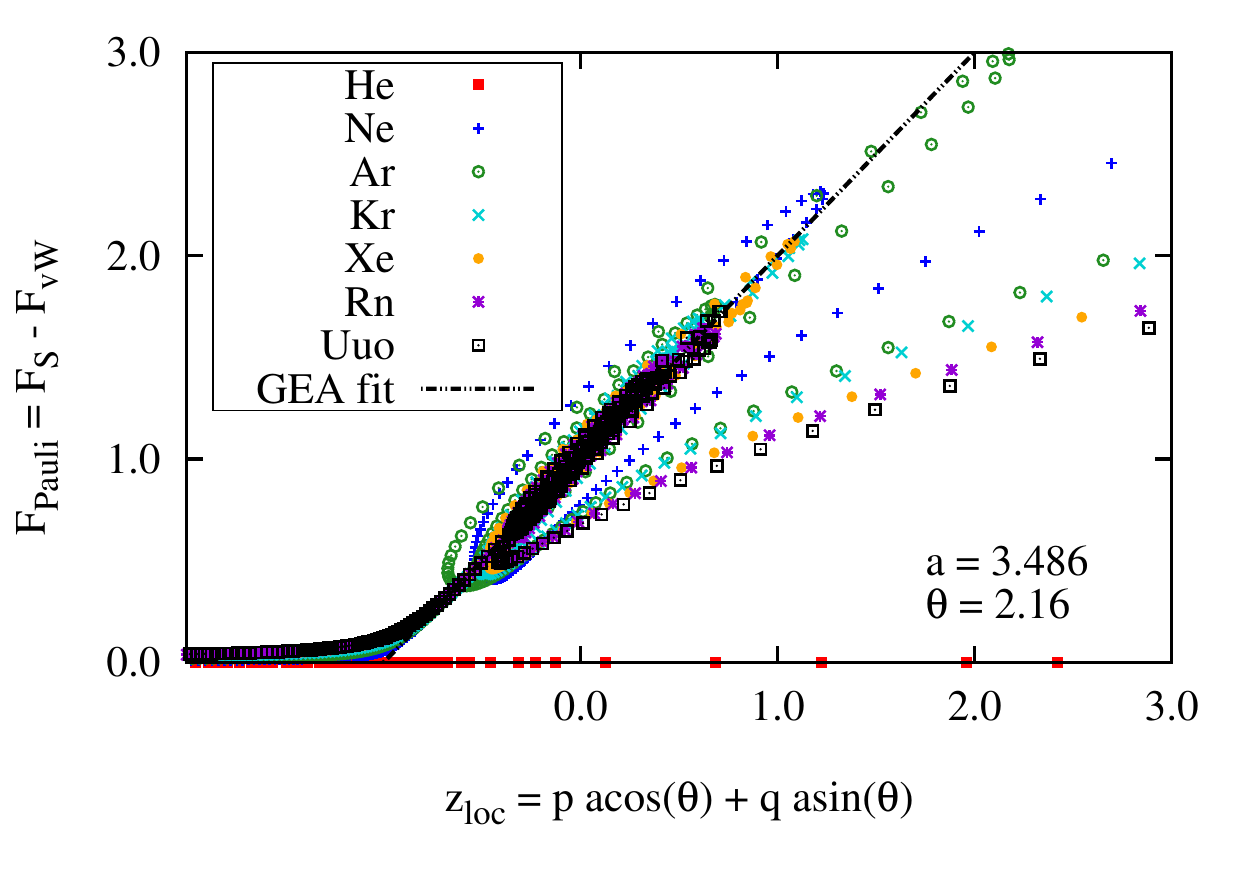}
      \caption{$F^{KS}_{Pauli}(r)$ plotted parametrically versus $\zfit(r)$ 
[Eq.~(\ref{eq:zfit})] for the noble gas atoms, including Helium and 
Unumoctium.  Dashed line gives the 
GEA fit $F^{GEAloc}_{Pauli} = 1 + \zfit$.  The values for $a$ and $\theta$ 
used to define $\zfit$ are those obtained by optimizing the fit for Uuo.
      \label{fig:pauliscatter}
}
\end{figure}

\begin{figure}[!htbp]\centering
      \includegraphics[width=0.9\linewidth,height=0.40\textheight,keepaspectratio]{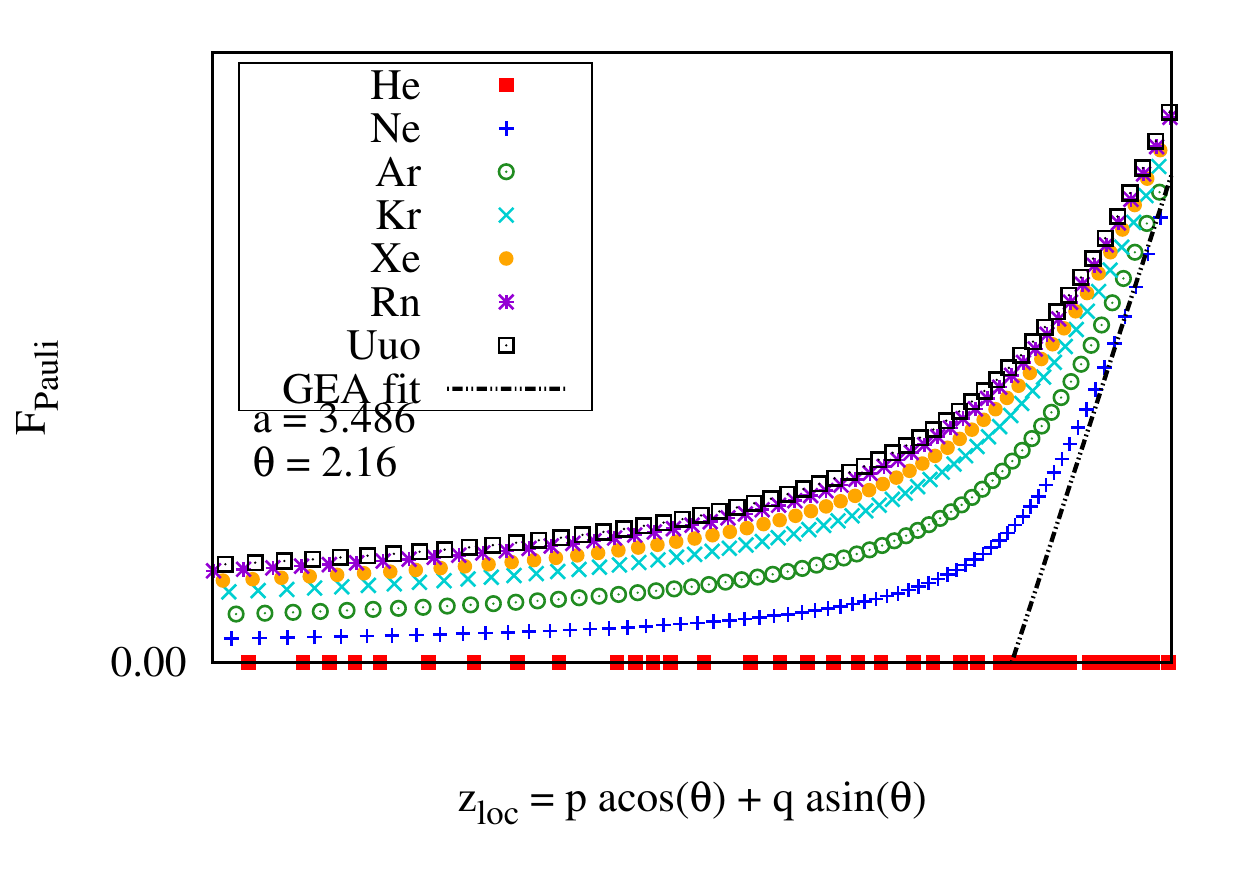}
      \caption{The same as Fig.~\ref{fig:pauliscatter}, but focusing on
    the near-nuclear regime where $q \rightarrow -\infty$. 
      \label{fig:pauliscatter_sht}
       }
\end{figure}

We have seen that the behavior of $F_{Pauli}$ for atoms projected upon the
parameter space defined by $p(\bfr)$ and $q(\bfr)$ is capable of a great
deal of simplification.  
Given the hypothesis that we might have 
a successful two parameter parametrization $F_{Pauli}[p(r),q(r)]$, we find
through Fig.~\ref{fig:3d} that we essentially only have a one-parameter space,
$F_{Pauli}[\zfit(r)]$, with $\zfit$ given by Eq.~(\ref{eq:zfit}).
The result is shown in Figs.~\ref{fig:pauliscatter} and~\ref{fig:pauliscatter_sht}.

These plots provide a wealth of detail that illuminate several key
features of the Kohn-Sham KED of atoms.  
Most important of all is the visualization of how the KED scales to high $Z$.  
A single shell system such 
as He has zero Pauli KED and is in this sense infinitely far from the
asymptotic limit.  But any two-shell system already captures
much of the sense of what happens at large $Z$, albeit
with obvious shell structure -- for example, $F_{Pauli}$ for Ne (blue crosses)
loops around but does not land on the GEA line.  
Here Be and Li, not shown, are worst cases, as one might expect, while Ne
is already fairly close to the limit.  
As more and more shells are added, $F_{Pauli}$ continues to 
loop around the large-$Z$ asymptote defined by the GE line, 
but in ever tighter loops that rapidly approach the asymptote.
There is a hint of curvature for Uuo
that might imply a fourth-order gradient correction but a very small
one, as is the case for the standard gradient expansion.

Second, we see the two regions that cannot be 
captured by Thomas-Fermi theory each demonstrate 
difficulties with the asymptotic model. 
First of all, the region $r \rightarrow \infty$ correlates with 
the loss of a well-defined single-valued function $F_{Pauli}(z)$.
That is, for any point in the core region of an atom corresponding
to some $\zfit$ and some value of $F_{Pauli}$, there will be a point in the 
asymptotic region with the same value of $\zfit$ but 
requiring a value of $F_{Pauli}$ up to 50\% smaller.  Moreover, every 
individual atom seems to require a unique form for $F_{Pauli}(z)$ in the
asymptotic region.  Though the tails seem to converge to some finite
 value as $Z$ increases, this convergence is also very slow.

This behavior may be an indication of the problem facing OFDFT in the 
asymptotic region discussed in Sec.~\ref{sec:limits}.
In this regime, the Pauli KED has a contribution from the HOMO shell
[Eq.~(\ref{eq:tauHOMO})] that depends upon
the angular momentum quantum number of the shell.  It therefore cannot 
be predicted from the total particle density alone.  
At the same time, it shoud  be noted that the worst behavior occurs
only for very large $r$. 
As seen in Fig.~\ref{fig:densAr}, the Pauli enhancement factor of the 
HOMO shell tends to be depressed relative to $p$ and $q$ and hovers around
its minimum value for a fair distance.  This is also seen as the clumping 
of a large number of grid points in Fig.~\ref{fig:pauliscatter} at the very
last local minimum in $F_{Pauli}$ before it trends off to $\infty$.  
The impressive near-universal form seen in Fig.~\ref{fig:3d} is a 
reflection of the gradual onset of non-universal behavior. 

A second difficulty occurs for the smallest radii, within the innermost
shell of each atom, as shown in Fig.~\ref{fig:pauliscatter_sht}.
Here the Pauli contribution to the KED is non-zero and measures the contribution
of $p$-orbitals to the KED.
Systems like He, Li and Be with no $p$-orbitals have exactly zero Pauli
KED in this limit, as seen for He in this plot.
For atoms with $p$ orbitals, the result depends sensitively on how many
shells are occupied, with the smallest $F_{Pauli}$ for Neon and the largest 
for Uuo.  There is a definite limiting case for 
infinite $Z$~\cite{DSFC15}, which is approached rather slowly.
The functional form of $F_{Pauli}$ for these systems is linear in $r$
at the nucleus -- the enhancement factor has a finite cusp.  This 
translates to a Pauli correction of the form $F_0(1 + A_0/\zfit)$ where
$F_0$ and $A_0$ necessarily depend upon the number of electrons.
Although this seems to be a very small effect, with $F_0$ on the order of
0.02 for the largest physical atoms, it occurs in a limit with extremely
high density and has a measurable impact upon integrated kinetic energies
as we shall see in the next section.

\subsection{Modified functionals for the KED}

We find two insights for developing OFDFT from the perspective of the 
local kinetic energy density.
First of all, rather than the canonical gradient expansion, which is 
derived to from an expression for the \textit{integrated} kinetic energy
of the slowly varying gas, we should start from the observed gradient
expansion for the local kinetic energy density.
In our mGGArev model, this is achieved by simply replacing the argument 
$z$ in Eq.~(\ref{eq:mggarev}) with $\zfit$ of Eq.~\ref{eq:zfit}.  
This produces a new 
family of possible functionals (mGGAloc$\alpha$) with different 
values of the parameter $\alpha$ that controls 
the rate at which the transition between gradient expansion and 
von~Weizs\"acker model occurs for strong electron localization.
Analogous corrections can be made for the mGGA. 


The second insight stems from the deviation of the KED from the gradient
expansion near the nucleus.
The nuclear region is a particular point of interest for models of the
local KED such as the mGGA 
and the related meta-GGA's we have constructed.
The transition to large negative values for the gradient expansion
correction that occurs in this region
breaks the basic constraint on the KED that $F_{Pauli} > 0$; in fact 
here $F^{GEA}_{Pauli} \rightarrow -\infty$.
This region is thus necessarily a probe of the 
transition from the slowly-varying electron gas characterized by the GE 
and the localized electron limit dominated by the von~Weizs\"acker 
KED.  
Exactly how the Kohn-Sham KED responds in this situation is a clue as to how 
to model this transition. 

\begin{figure}[!htbp]\centering
      
    \includegraphics[width=0.9\linewidth]{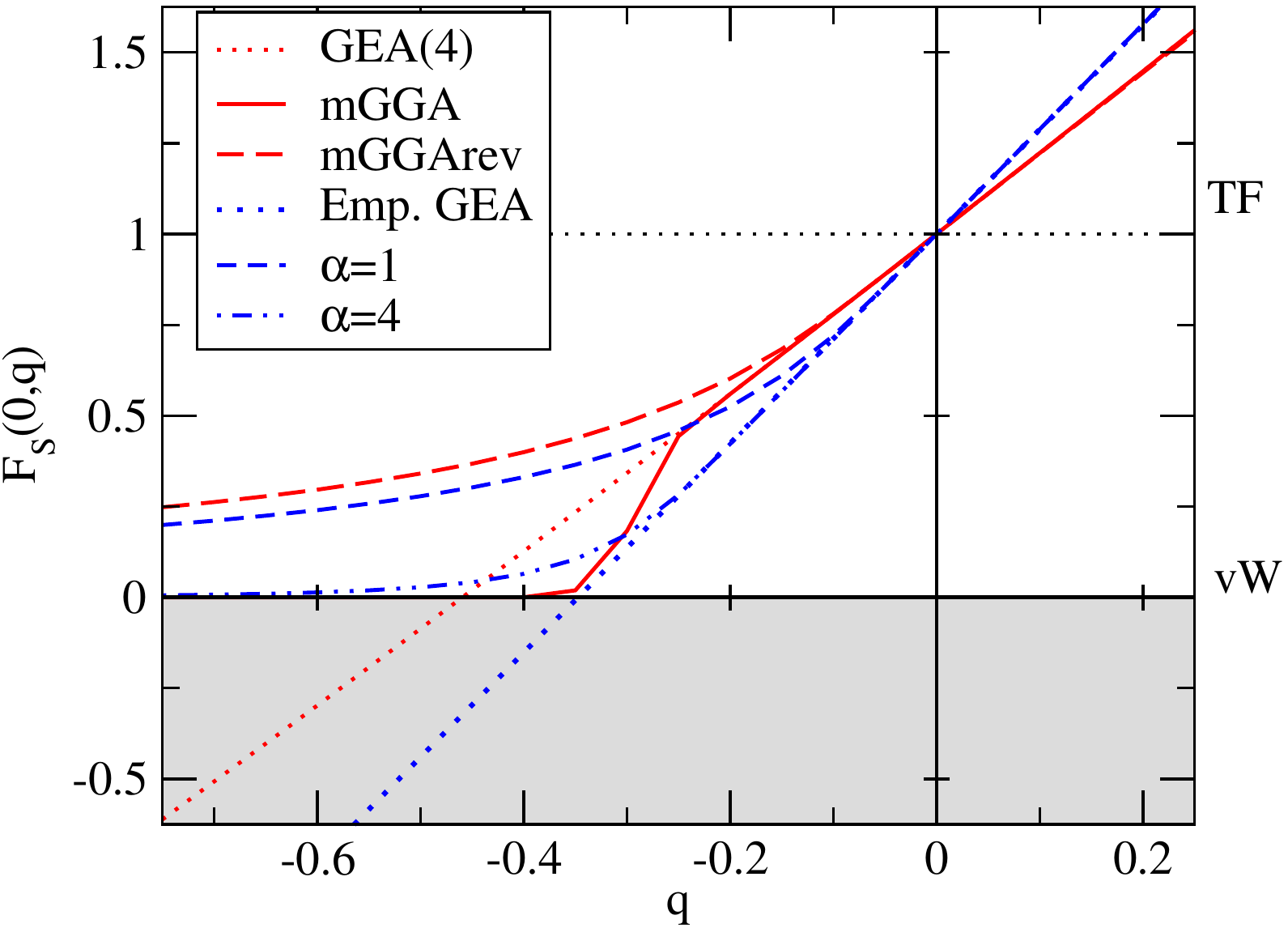}
      \caption{The Pauli enhancement factor for the fourth-order GEA (dotted),
    meta-GGA's based upon it (mGGA, mGGArev), 
    and for the empirically fit second-order GEA (widely-spaced dotted) and 
    variants of the mGGArev built upon it, using $\alpha=4$ 
    and 1 in Eq.~\ref{eq:mggarev}. 
    Shown versus $q$ for $p=0$, approximating 
    the conditions near the atomic nucleus. Grey area shows region forbidden
    by the von~Weizs\"acker bound.
    \label{fig:fmgga0qsmall}
       }
\end{figure}

The impacts of the varying strategies for doing this are shown
in Fig.~\ref{fig:fmgga0qsmall}.  This plots enhancement factors $F_S$
for the special case of zero density gradient versus the Laplacian-based
variable $q$.  This limit is a fair approximation of the nuclear region,
where $p$ is small ($<0.2$) and nearly constant while $q$ tends to $-\infty$,
as shown in Fig.~\ref{fig:3regions}.  In this case, $\tauvW \!=\! 0$ so 
that the lower bound it imposes is easy to visualize: 
$F_S \!=\! F_{Pauli} \!>\! 0$.

The canonical gradient expansion is shown to fourth order (dots), 
very nearly a straight line in $q$ passing through the Thomas-Fermi 
limit $F_S\!=\!1$ at $q\!=\!0$.  It very quickly goes below zero for negative 
$q$.  
The empirical local GEA (wider-spaced dots) 
exaggerates this behavior, given its steeper
slope in $q$, evident in Eq.~(\ref{eq:TGEAloc}). 
The mGGA imposes $F_{Pauli} > 0$ by a sharp cutoff that interpolates between
GEA and von~Weizs\"acker functionals in such a way as to be identically zero 
for negative $q$ beyond the GEA crossover point.  
The mGGArev [Eq.~(\ref{eq:mggarev}), with $\alpha\!=\!1$] is shown as 
long-dashed line.  This enforces $F_{Pauli} \!>\! F^{GEA}_{Pauli}$ which is 
beneficial for molecular bonding~\cite{CSK16}.
The short-dashed and dot-dashed lines show the mGGAloc with $\alpha\!=\!1$ and 
$\alpha\!=\!4$, which adhere to the local GEA outside the transition region.

Two points may be learned from this comparison. First of all, the functional 
form of the mGGArev is closer to reality than that of the mGGA.
As seen in Fig.~\ref{fig:pauliscatter_sht} 
the KS KED tapers off like the blade 
of a hockey stick, as $q$ and thus $z \rightarrow -\infty$, 
and certainly lacks the mGGA's abrupt transition to zero.
In that sense, the hypothesis upon which the mGGArev is based~\cite{CSK16} -- 
that $F^{KS} > F^{GEA}$ as $q \rightarrow -\infty$ --
does hold here, as long as one uses the empirical \textit{local} GEA, and not 
the canonical GEA. 

However, as we shall see next, the mGGA is highly
accurate for the total kinetic energy of atoms, while the mGGArev and its
relation the mGGAloc1 give large overestimates.  While having the
correct qualitative shape, they both overestimate the contribution
to the integrated KE from this region.  Only the mGGAloc4 approaches the
quality of the mGGA.  The mGGA's success thus seems to be from a clever
weaving from the wrong gradient expansion limit, to the 
wrong approach to the von~Weizs\"acker limit in such a way as to cancel
out the errors from each region.
Getting a better local KED does not guarantee a better kinetic energy,
thus meriting serious attention to the integrated quantity. 

\subsection{Integrated Kinetic Energy}
\begin{figure*}[!htbp]\centering
    \subfigure[]{
    \label{fig:TvsZ_models}
    \includegraphics[width=0.47\linewidth,height=0.25\textheight,keepaspectratio]{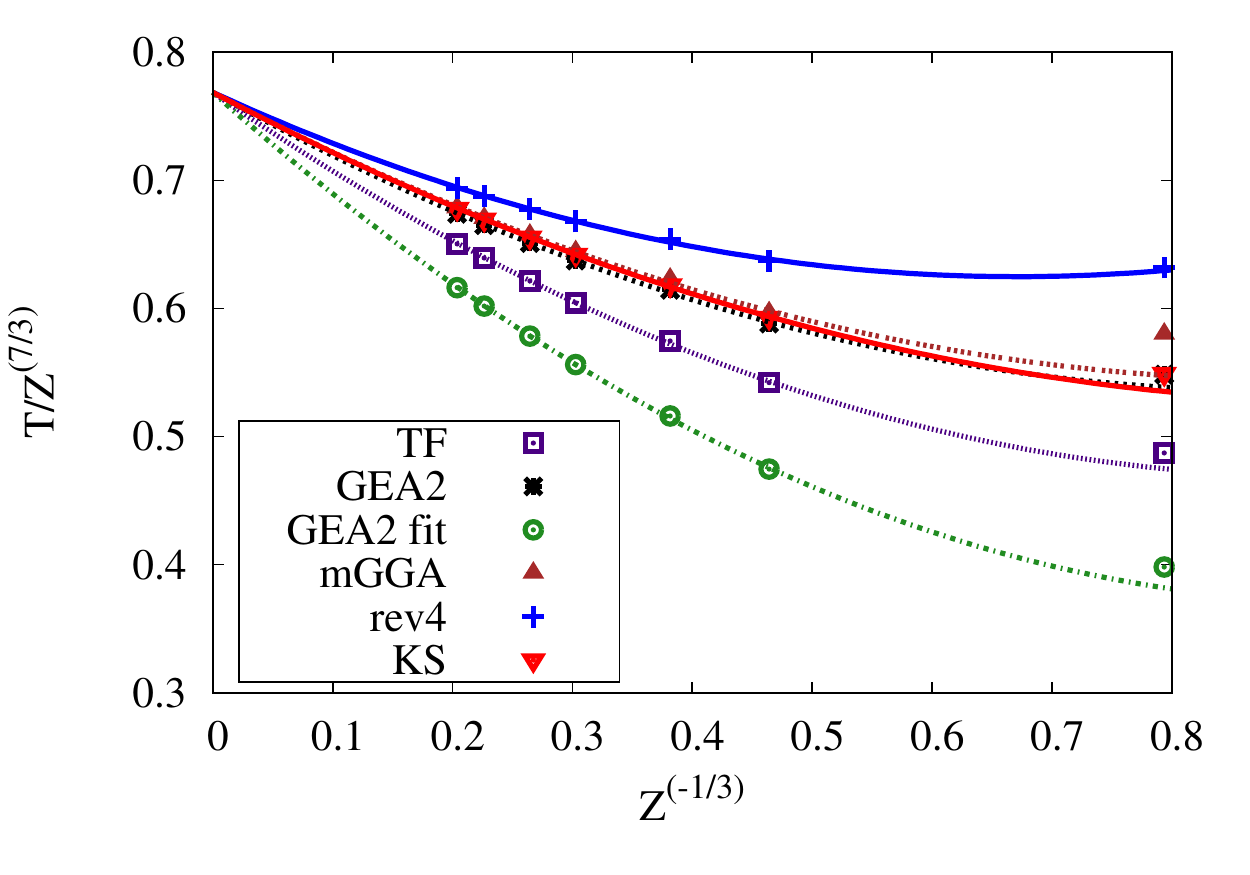}
    }
    \subfigure[]{
    \label{fig:TvsZ_models_zoom}
    \includegraphics[width=0.47\linewidth,height=0.25\textheight,keepaspectratio]{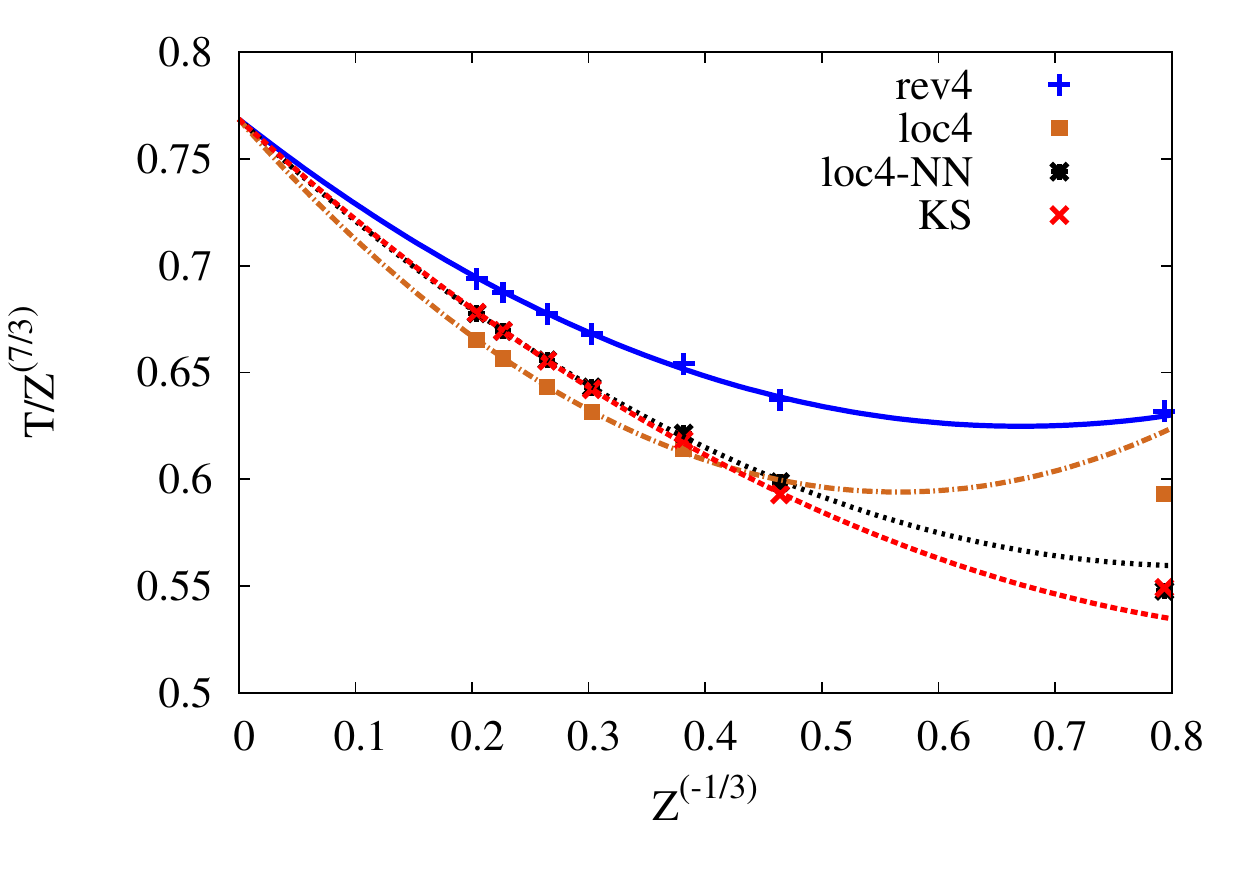}
    }
    \caption{(a) $T/Z^{7/3}$ versus $Z^{-1/3}$ for standard kinetic energy 
models discussed in the paper. GEA2-fit is the second-order GEA using empirical 
parameters of Eqs.~(\ref{eq:geafit}) and~(\ref{eq:zfit}). (b) the same,
demonstrating the effect of using the empirical local GE in constructing
OFDFT models.
\label{fig:TvsZ}
}
\end{figure*}

Figures~\ref{fig:TvsZ_models} and~(b) show the 
integrated kinetic energy of the noble-gas atoms for many of the OFDFT 
models discussed in this paper, scaled by the Thomas-Fermi 
scaling factor $Z^{7/3}$ and plotted as a function of $Z^{-1/3}$.  
As discussed in Sec.~\ref{sec:theory}, the kinetic
energy can be expressed as an expansion in powers of $Z^{-1/3}$, with the 
infinite-$Z$ limit of 0.768745$Z^{7/3}$ predicted by Thomas-Fermi theory.
Also shown is a fit of the trend with $Z$ for each functional to the 
asymptotic form [Eq.~(\ref{eq:largeZ})].
The Thomas-Fermi limit is assumed for each case and the next 
two coefficients $B$ and $C$ are determined by linear regression over the noble
gases excluding He.  
The fit coefficients and errors are shown in Table~\ref{table:TvsZ}.

\begin{table}[!htpb]
        \caption{Least squares fit parameters for the $Z$ expansion 
        [Eq.~(\ref{eq:largeZ})] of the noble gases for various OFDFT 
        models of the kinetic energy.  The Thomas-Fermi 
        limit $A \!=\! 0.7687$ is assumed.}
   \begin{center}
       \begin{tabular}{l|c|c}
           \hline \hline
            Model & B & C \\ \hline
            Accepted   & $-1/2$         & 0.2699  \\ \hline
            KS/LDA   & -0.4943(43) & 0.252(11) \\
            TF       & -0.649(7)  & 0.351(19) \\
            GEA     & -0.522(8)  & 0.292(20) \\
            APBEK    & -0.489(8)  & 0.241(21) \\
            VT84F    &  0.116(20) & 0.72(8) \\
            mGGA     & -0.493(9)  & 0.270(23) \\
            mGGArev4 & -0.429(7)  & 0.320(20) \\
            GEAloc   & -0.834(6)  & 0.437(16) \\
            mGGAloc4 & -0.618(5)  & 0.546(13) \\
            fit4-NN  &  -0.4933(31) & 0.273(5) \\ 
            \hline\hline
        \end{tabular}
  

        \label{table:TvsZ}
    \end{center}
\end{table}

The slight disagreement between the theoretical and calculated asymptotic 
coefficients for the KS/LDA kinetic energy in Table~\ref{table:TvsZ} are 
within two standard deviations for the fit and thus seem reasonable. 
The errors due to the use of the LDA rather than exact KS density are
probably much smaller. 

Beyond this, it is possible to distinguish two classes of functionals.
The canonical GEA obtained from the slowly-varying electron gas is already
exceptionally close to the KS value and more sophisticated models like the 
mGGA struggle to improve upon or even do as well as it over all $Z$.
Nevertheless,
both it and the APBEK~\cite{APBE} are constructed in part through a fit 
to the large-$Z$
limit.  As a result both have excellent estimates of the asymptotic 
coefficients $B$ and $C$ and are nearly flawless for larger $Z$.

\begin{figure}
\includegraphics[width=0.9\linewidth]{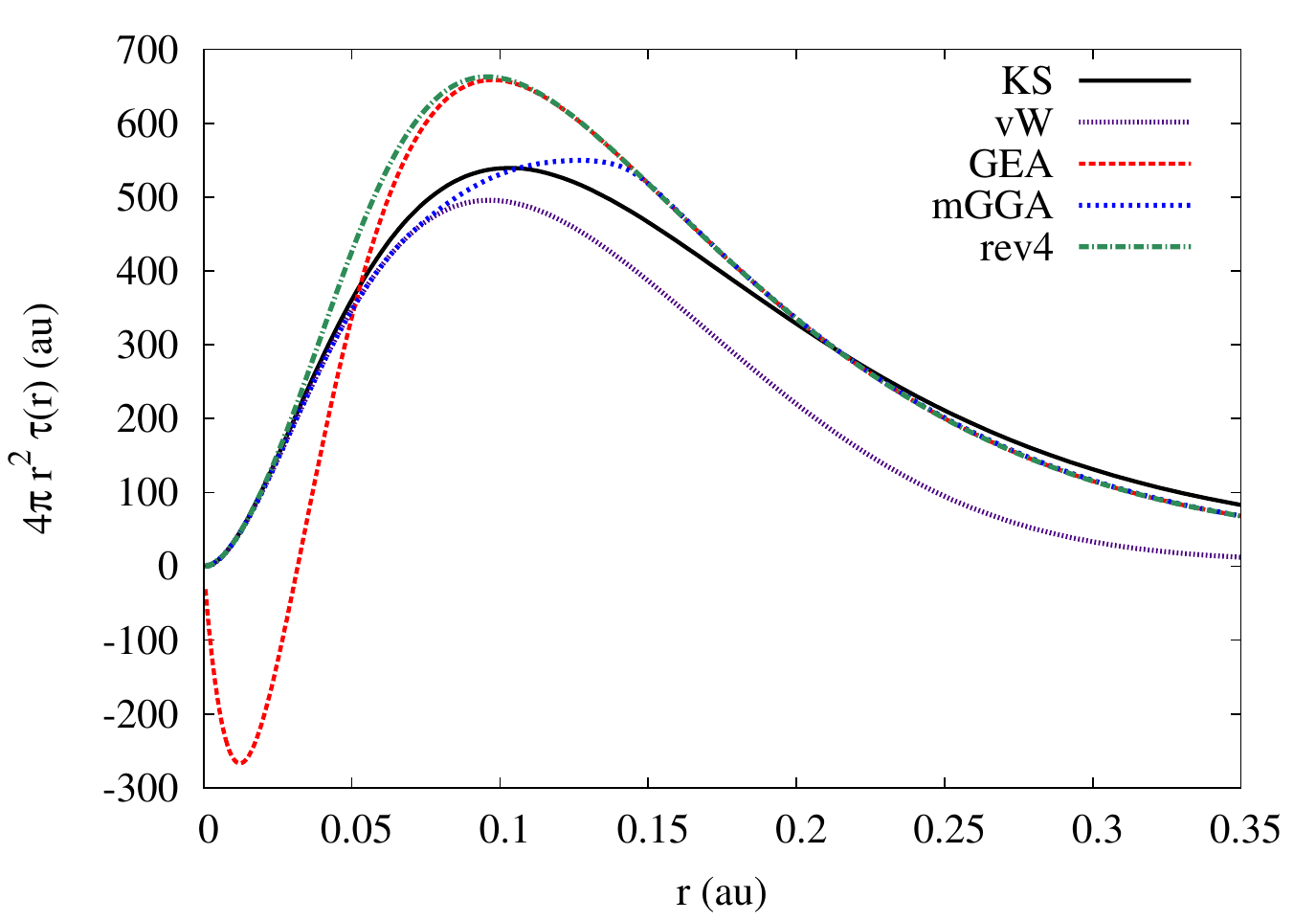}
\caption{
\label{fig:tauNe}
(colour online)
Kinetic energy radial densities in the 1s shell of the Ne atom.}
\end{figure}

On the other hand,
the mGGArev4, [Eq.~(\ref{eq:mggarev}) with $\alpha\!=\!4$, 
labelled rev4 on the plot] is a serious regression, and the VT84F, 
whose asymptotic coefficients are shown in Table~\ref{table:TvsZ}, is worse.  
These have been constructed with constraint
choices that emphasize the von Weiz\"acker lower bound on the KED.  In
 the mGGArev4 \textit{and} in the VT84F, this is done by imposing the 
implicit constraint that $\tau > \max(\tauvW,\tauGEA)$, the former by choice
and the latter by necessity given the restricted flexibility of the GGA
form.  This leads to an overestimate of total energy, because the GEA
is significantly less than the von~Weizs\"acker KED especially near the 
nucleus.  Removing this unphysical behavior must cause a net increase in the 
total kinetic energy, whereas the GEA is already almost perfectly accurate.
In contrast,
the mGGA interpolates between slowly-varying and von~Weizs\"acker
limit with a function that incorrectly obeys $\tauvW < \tau < \tauGEA$ --
thus taking advantage of a natural cancellation of errors.
Both of these effects are clearly seen in Fig.~\ref{fig:tauNe}, which 
shows the radial KE density of the 1s shell of Neon.  The GEA (dotted line)
has a large negative error at the cusp, but an equally large error at 
the peak of the shell.  In transitioning from the GEA to the vW, the
mGGA preserves this error cancellation.  The mGGArev4 (dot-dashed line) 
fixes the error near the cusp but its constraint choice prevents
it from fixing the error at the shell peak.

The final key to the story is 
the impact of the empirical local GEA we find for the KED.  The 
impact of its deviation from the standard GEA is to lower the local KED 
with respect to it everywhere in the system.  This produces a total KE 
that is much less even than the TF energy, as seen in Fig.~\ref{fig:TvsZ}.
At the same time, this lowering of KED works naturally with the raising
of energy caused by the imposition of the constraint 
$\tau > \tauvW$ near the nucleus and the further constraint 
$\tau > \tau_{GEAloc}$ that we have observed throughout the 1s shell.
The effect of combining this constraint with the local gradient expansion
is shown in Fig.~\ref{fig:TvsZ_models_zoom}.  
While using the canonical 
gradient expansion with these constraints leads to the serious overestimate
of the mGGArev4, the combination of the right form
of local gradient expansion with this constraint (labelled loc4) 
combine to almost cancel this error.

Unfortunately the overall quality of the asymptotic trend of the
mGGAloc4 with $Z$ is poor, as shown especially in Table~\ref{table:TvsZ}.  
This is the downside of the good cancellation of errors seen in the GEA:
removal of one error-causing effect leads to poorer results unless
the companion effect causing the cancellation is treated 
equally well.
The problem here is the
failure to account for the Pauli contribution from $p$ orbitals in the 
near-nuclear region, which has a measurable effect on the quality of the 
answer.  Thus
a model for this effect is necessary, if only to understand the physics
of the atom.  

%

\subsection{Empirical model of near-nucleus region}

In the previous section we have taken as a reference model the revised
mGGA of Eq.~(\ref{eq:mggarev}) with a transition parameter of $\alpha=4$. 
This is a reasonable choice -- it ensures that both the 
Pauli contribution to the KED and its potential 
$\delta \tau_{Pauli}(\bfr) / \delta n(\bfr)$ are zero near the nucleus.
This ensures that for systems like He, for which there is no Pauli 
KED, or for small $Z$ in general, that the near-nuclear region at least
is handled reasonably.  (It is improbable that a functional
based upon the slowly-varying electron gas can produce zero $\tau_{Pauli}$ 
\textit{everywhere}.)
However, this choice of interpolating factor does not account for the non-zero 
contribution by $p$-orbitals to the Pauli KED at the nucleus. 
Unfortunately, we have seen 
(Table~\ref{table:TvsZ} and Fig.~\ref{fig:TvsZ_models_zoom})
that our best empirical fit for the core and
asymptotic regions gives a poor estimate for integrated KE's of atoms.
This indicates that the error in ignoring the 
Pauli contribution to the KED near the nucleus is a measurable effect.  
Though the Pauli enhancement factor in this region is 
small (Fig.~\ref{fig:pauliscatter_sht}), it results in a significant 
contribution to the KE given the enormous densities for 
large-$Z$ atoms.  And unfortunately, we need a correction that is different
for every row of the periodic table, each of which adds a new $p$ orbital  
to the system and an additional contribution to the Pauli KED. 
Thus a correction to the von~Weizs\"acker KED is required for this
region that is somehow dependent upon the electron number $N$.  

As a first step in this direction, we build upon the $N$-dependent
model developed by Acharya et al.~\cite{Acharya} 
Their work noted that an excellent model of the KED
for atoms could be obtained by first taking a slowly-varying model of the 
KED such as the TF or GEA model for all shells but the innermost K shell.  
Then, for the K shell, the model is replaced by the von Weizs\"acker KED:
\be
    \tau[n] = \tau_0[n] - \tau_0[n_K] + \tauvW[n_K]
\ee
with $\tau_0$ the KED of the initial slowly-varying model and $n_K$
the density of the K-shell.  Note that at the nucleus
this model essentially restricts $\tau_0$ to the description of the small
Pauli contribution to the KED due to $p$ orbitals,
and assumes that $\tauvW$ contributes negligibly elsewhere.
With reasonable assumptions about the nature of the K shell density $n_K$,
one gets an $N$-dependent model for the KE:
\be
    T[n] = \frac{T_{vW}[n] + T_0[n]}{1 + c/N^{1/3}}.
\ee

A very similar approach has recently been proposed~\cite{CFDS16} which uses the 
KED of the K-shell as a basic variable for building an OFDFT and extending
the analysis to treat the exchange contribution from this shell.
It provides excellent predictions of exchange and kinetic energy densities 
near the nucleus, suggesting that the careful treatment of the K-shell 
density is the key to modeling the KED in this region.
We will take another tack to this issue, by determining 
an $N$-dependent correction to the mGGArev functional that reproduces 
the important features of the Acharya KED in the near-nuclear regime and
recovers the asymptotic scaling of the KE of atoms to large $Z$. 
We do so by modifying the mGGArev interpolation function $I(z)$, using
$z\! =\! \zfit$, to 
\be
      I_{NN}(\zfit,N) = \left \{ 1 - 
                   \exp{\left [-\beta^\alpha(N)/|\zfit|^\alpha \right] }
                   H(-\zfit) \right \}^{1/\alpha}
      \label{eq:mggaNN}
\ee
where
\be
    \label{eq:oursNN}
    \beta(N)=A_{NN}+\frac{B_{NN}}{N^{1/3}}.
\ee
Expanding about the near-nuclear limit $\zfit \to -\infty$ 
we find 
\be
     \lim_{z \rightarrow -\infty} F_S^{mGGAnn} = \FvW + 1 - \beta.
\ee

Essentially, the correction contributes a non-zero component to Pauli
KED in the near-nuclear region with the same scaling in $N$ as the empirical
Acharya correction.
By adjusting the constants 
$A_{NN}$ and $B_{NN}$, our functional can be empirically fit to the $Z$ 
scaling behavior of the KS KE for large $Z$ atoms.  
Our original model is recovered with $A_{NN}\!=\!1, B_{NN}\!=\!0$.
Values of $A_{NN}\!\approx \!0.77$ and 
$B_{NN}\!\approx \!0.50$ give a nearly ideal fit to the Kohn-Sham kinetic
energy as seen in Fig.~\ref{fig:TvsZ_models_zoom}.  These are remarkably
close to the large-$Z$ expansion 
parameters of Eq.~(\ref{eq:largeZ}), although we have no evidence 
that this is more than a coincidence.  

\begin{figure*}[!htbp]\centering
    \subfigure[]{
    \label{fig:tauAr_diff}
    \includegraphics[width=0.47\linewidth]{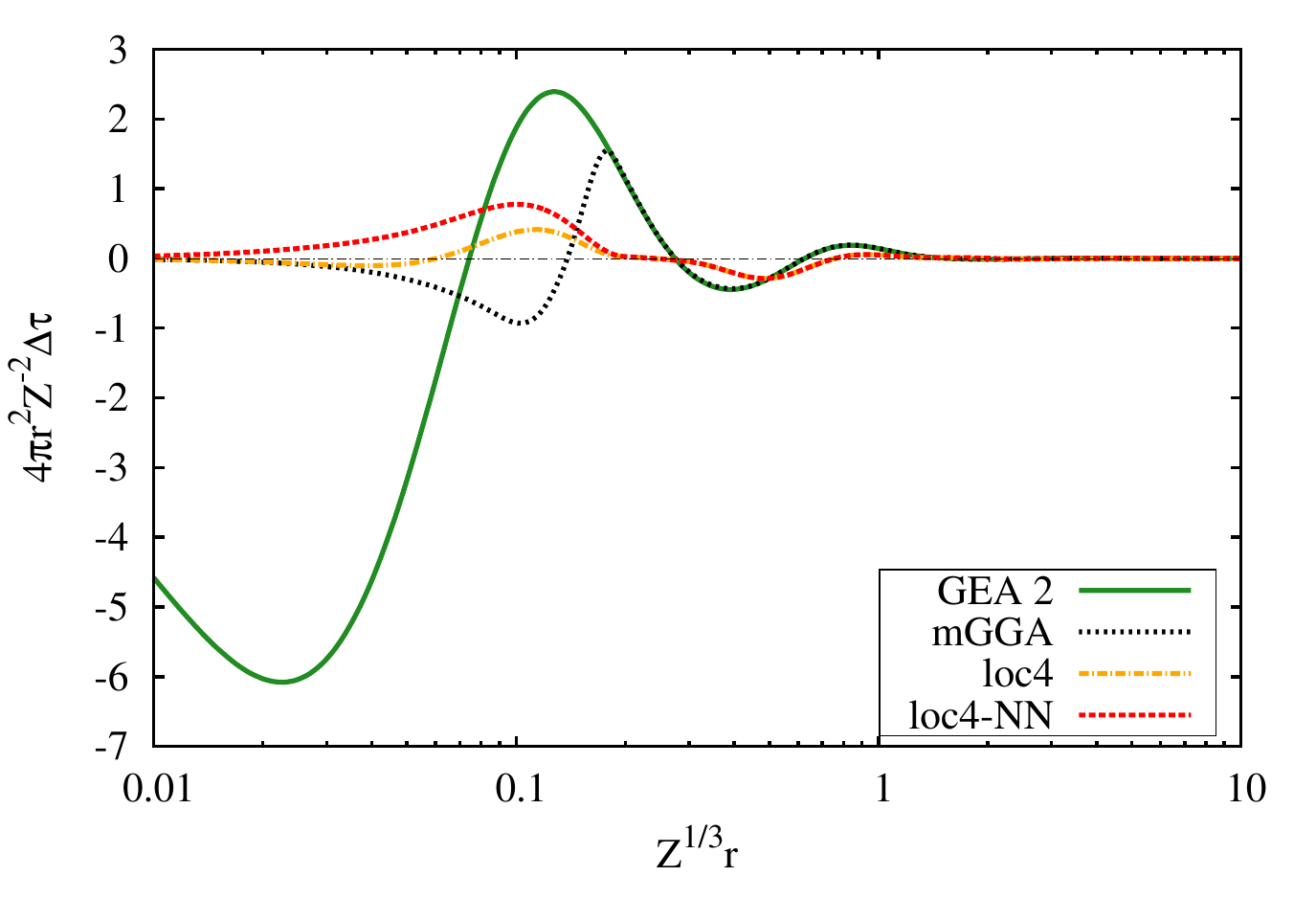}
    }
    \subfigure[]{
    \label{fig:tauUuo_diff}
    \includegraphics[width=0.47\linewidth]{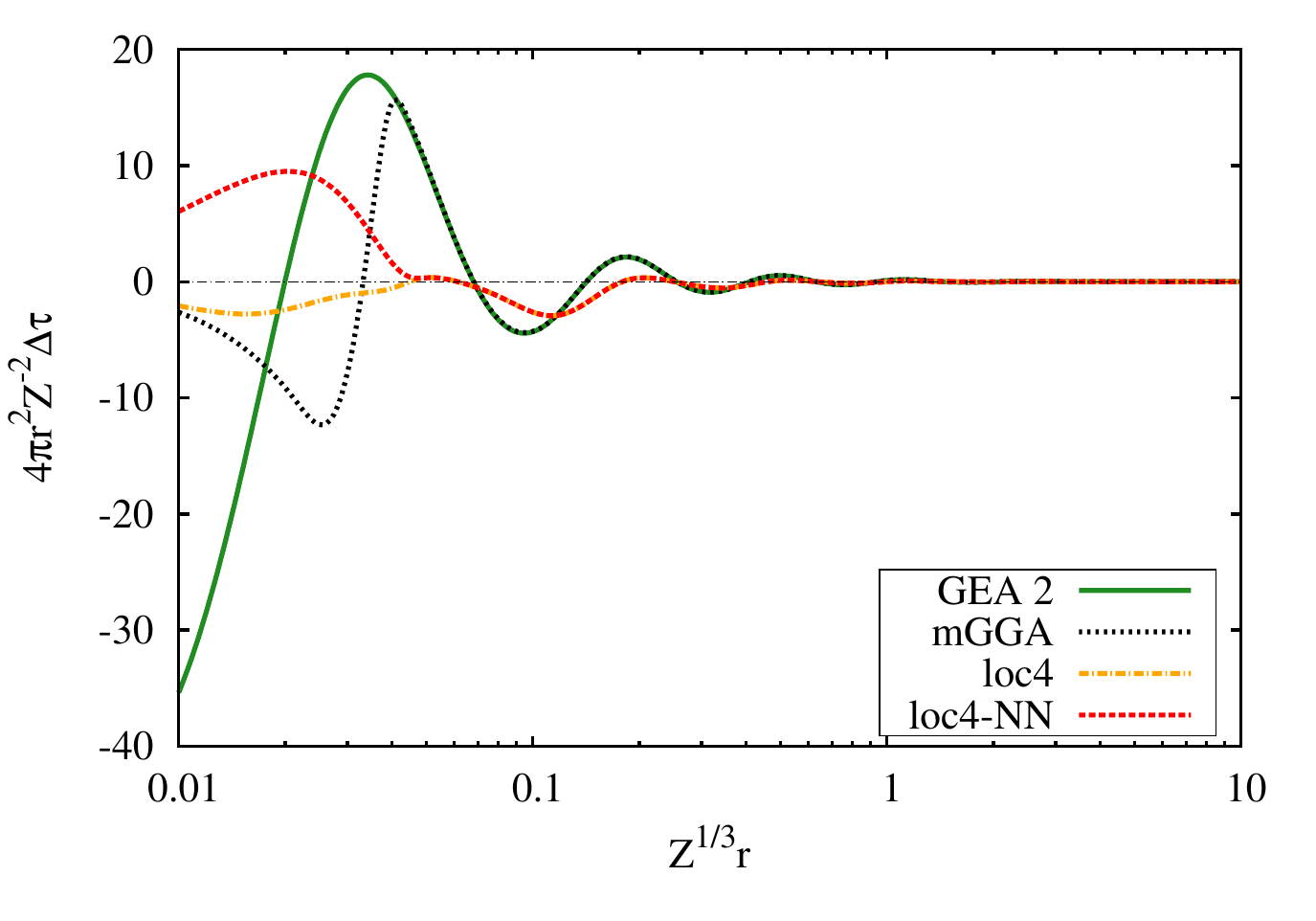}
    }
    \caption{(a) Error in the scaled radial KED of Argon 
    $4\pi \bfr^2[\tau_{model}(\bfr)-\tau_{KS}(\bfr)]/Z^2$ 
    versus scaled radius for several KED models.
     (b) The same, for unumoctium ($Z\!=\!118$).
     \label{fig:tau_diff}
     }
\end{figure*}

Nevertheless, these values 
are poor predictors of the actual KED at the nucleus -- while the
actual value of $F_{Pauli}(r\!=\!0) \sim 0.022$ at the nucleus for Rn, 
our correction predicts a value six times larger.  
This is indicated by 
error introduced into the KED as $r \!\to\! 0$, as seen for Argon
in Fig.~\ref{fig:tauAr_diff} and Uuo in Fig.~\ref{fig:tauUuo_diff}.  
The excellent KE's are caused by
successful cancellation of errors between those of the near-nuclear regime  
and that accumulated across the rest of the atom.  Interestingly,
the need is to make the fit in the near-nuclear region worse compared
to the non-$N$-dependent mGGAloc model.  By comparing 
Fig.~\ref{fig:tauAr_diff} 
to Fig.~\ref{fig:densAr}, we find the second largest
source of error for the mGGAloc4 (fit-4 in the plot) comes in the transition 
between shells, where $F_{Pauli}$ has
a local maximum.  This error is already outside the nuclear cusp region
and in that of oscillatory behavior of $F_{Pauli}$ about the 
gradient expansion asymptote as seen in Fig.~\ref{fig:pauliscatter}.  
As $Z$ increases, the magnitude of the error increases, and more shells seem
to be involved,
but its contribution to the total KE decreases, as the region of error moves
farther from that of peak radial charge density at $Z^{1/3}r \sim 1$.
Fig.~\ref{fig:tau_diff} tells roughly the same story for the mGGA, and 
the cancellation of error in that model,
but with generally larger amplitude oscillations. 

\section{Discussion and Conclusions\label{sec:conclusion}}
We have analyzed scaling trends in the positive-definite 
Kohn-Sham kinetic energy density 
over the periodic table of atoms. 
We have concentrated our attention to the transition to the large-$Z$ limit,
in order to characterize the
diminishing size of the corrections to the Thomas-Fermi limit as $Z$ increases.
Second-order density derivatives $\lapln$ and $\gradnsq$
expressed in scale-invariant form provide a intuitively useful and 
nearly complete visual description of the atom and particularly, the trends 
with $Z$
of different local regions of the atom -- nucleus, core, and valence shell.
The pair thus should be a useful basis for constructing orbital-free maps 
of local quantities such as the kinetic energy density or the energy
densities associated with the exchange and correlation holes.

In fact, we find that over much of the atom,
corresponding roughly to the regime of validity of the TF model in
the infinite-$Z$ limit, the Kohn-Sham KED is exceptionally well
fit by a simple second-order gradient expansion.  
For low $Z$ 
deviations from this asymptotic trend, caused by shell structure,
naturally oscillate about it and gradually reduce as $Z$ increases.
At large $Z$, the local GE model 
becomes nearly exact, and independent of column.  
This suggests that the gradient expansion is the fundamental semilocal
density functional correction to the TF limit, but with the significant
caveat that the local gradient expansion for the KED is not the global one
for the KE.
In fact it is qualitatitively different -- the correction to the 
integrated Thomas-Fermi KE obtained
from the local gradient expansion is the opposite sign of the normal case.  
Thus, we cannot say that if $T_{KS} = \int \tau_{model}(\bfr) d^3r$ then 
$\tau_{KS}(\bfr) = \tau_{model}(\bfr)$, or vice-versa.
Note that this is not simply an issue of choice of
``gauge", where one might compare two KED's defined in alternate ways that
integrate to the same value.  In this paper, only the unique
positive-definite gauge is used.

Rather the problem is fundamental -- the relative success of Kirzhnits GE 
is not 
because of the accuracy of the underlying local functional $\tau[n(\bfr)]$ 
because this breaks the lower bound $\tau \!>\! \tauvW$.  It rather captures 
a cancellation of errors in the integral of $\tau$ -- the breaking of the 
von~Weizs\"acker lower bound near the nucleus being 
compensated by an overestimate of the local gradient correction elsewhere.
This points to the much greater difficulty in modeling the local versus the 
global quantity, as the former requires modeling from point to 
point and is thus much less amenable to beneficial error cancellation.
At the same time these results confirm, qualitatively if not quantitatively,
the gradient expansion analysis of Ref.~\cite{LAM14} for the Airy gas,
a model designed asymptotically to represent a system that is all surface.
Together these two asymptotic limits strongly suggest that the Kirzhnits
gradient expansion should not be used in an application (presumably
including bonding) that would depend sensitively on the local kinetic 
energy density.

It is not surprising to find that the greatest difficulties in removing
this point-to-point error using the second-order gradient quanitities
$p$ and $q$ are the two limits in which Thomas-Fermi theory fails.
The asymptotic limit far from the atom is problematic because the Pauli
kinetic energy deviates from being a single-valued function of these 
variables.
This seems to be related to the dependence of $\tauKS$ on the angular
momentum quantum number $l_{HOMO}$ of the HOMO orbital~\cite{DSFC15},
something that is not predictable with only $\lapln$ or $\gradnsq$.
The use of higher-order derivatives might help in this case~\cite{dSC15}.
The near-nuclear region dominated by the cusp in the density is also
difficult because of the sensitive dependence of the Pauli KED on the 
number of electrons occupying $p$-orbitals in the system.  This might be 
crudely approximated with the reduced density gradient $p$, which also
shows a weak dependence on $N$,
but the recent nonlocal approach of Ref.~\cite{CFDS16} 
should be more robust.

In all then, it is not surprising that a simple fix to our OFDFT meta-GGA
models, 
replacing the global gradient expansion with the empirical local one we 
find here, fails to produce good total kinetic energies for the atoms.  
While they can hit the ballpark of KS energies, they do not compare 
favorably even to the lowest level conventional gradient correction. 
Rather our findings should help to develop OFDFT models that much more 
accurately model the KED in the bulk of the atom than prior models. 
In this, the fact that we can limit the functional to a gradient expansion 
and not a GGA helps a lot -- a gradient expansion has a well behaved 
Pauli potential that neither breaks known constraints nor generates
unphysical oscillatory behavior.

At the same time, we can reproduce the integrated KE of atoms with 
excellent accuracy
given a fit to a simple $N$-dependent modification of our orbital-free
model.
A density-functional theory that depends upon the number of electrons
$N$ may be less than satisfactory from an \textit{a priori}
standpoint.
More to the point perhaps is that this close fit is achieved by introducing,
not reducing, error into the KED at the nucleus in order to cancel out the net
error from inner shells.  A connection to semiclassical theory may
explain this.  
In a paper deriving the Scott correction $B\!=\!-1/2$ 
to the Thomas-Fermi KE~\cite{Schwinger80}, Schwinger 
noted that the correction came not just from the cusp region where the 
Thomas-Fermi density diverges, but also from quantum
oscillations in the inner shells --
those with peaks at radii $r_{peak} \ll Z^{1/3}a_B$.  In our situation, for
even the largest system, Uuo, we see not only large errors at the nucleus,
but in the quantum oscillations about the gradient expansion that damp out 
only gradually.  
We believe that for $Z\to\infty$ these oscillations will remain
large for any atom, but extending only over a fraction of the inner shells,
becoming negligible relative to the total energy.
The point is that a successful model of the KED for atoms will have to account
for both the unusual Pauli energy density \textit{at} the nucleus and 
for large quantum oscillations in the nearby shells.  
The progress made to handle the former in Ref.~\cite{CFDS16} 
will need to be matched by improvement in the
latter; these might be made by a fourth-order gradient correction. 

A final issue is whether the use of a negative gradient correction in the 
gradient expansion helps to improve binding energies predictions for 
molecules, or perhaps makes them worse.  
This issue is currently being explored.  Preliminary data 
for the AE6 test set show that the use of a mGGAloc using the atomic 
local GEA rather than a mGGArev using the conventional GEA does improve 
binding energies consistently.  At the same time, the indication
is that this improvement is nowhere near enough 
to make OFDFT competitive with Kohn-Sham methods.  
However, it would be interesting to explore the effect of the use of a 
negative gradient expansion coefficient in a GGA. If the best
performer on the test set, the VT84F,
showed a similar improvement in binding energy we see for our meta-GGA's, it
should come within the ballpark of the LDA in performance.
Our findings thus
should make a contribution, if not a decisive one, towards solving the 
challenge of the orbital-free prediction of covalent bonding. 


\section*{Acknowledgments} 
A.C.C would like to thank Kieron Burke
and Sam Trickey for useful discussions.  


%


\begin{thebibliography}{71}%
\makeatletter
\providecommand \@ifxundefined [1]{%
 \@ifx{#1\undefined}
}%
\providecommand \@ifnum [1]{%
 \ifnum #1\expandafter \@firstoftwo
 \else \expandafter \@secondoftwo
 \fi
}%
\providecommand \@ifx [1]{%
 \ifx #1\expandafter \@firstoftwo
 \else \expandafter \@secondoftwo
 \fi
}%
\providecommand \natexlab [1]{#1}%
\providecommand \enquote  [1]{``#1''}%
\providecommand \bibnamefont  [1]{#1}%
\providecommand \bibfnamefont [1]{#1}%
\providecommand \citenamefont [1]{#1}%
\providecommand \href@noop [0]{\@secondoftwo}%
\providecommand \href [0]{\begingroup \@sanitize@url \@href}%
\providecommand \@href[1]{\@@startlink{#1}\@@href}%
\providecommand \@@href[1]{\endgroup#1\@@endlink}%
\providecommand \@sanitize@url [0]{\catcode `\\12\catcode `\$12\catcode
  `\&12\catcode `\#12\catcode `\^12\catcode `\_12\catcode `\%12\relax}%
\providecommand \@@startlink[1]{}%
\providecommand \@@endlink[0]{}%
\providecommand \url  [0]{\begingroup\@sanitize@url \@url }%
\providecommand \@url [1]{\endgroup\@href {#1}{\urlprefix }}%
\providecommand \urlprefix  [0]{URL }%
\providecommand \Eprint [0]{\href }%
\providecommand \doibase [0]{http://dx.doi.org/}%
\providecommand \selectlanguage [0]{\@gobble}%
\providecommand \bibinfo  [0]{\@secondoftwo}%
\providecommand \bibfield  [0]{\@secondoftwo}%
\providecommand \translation [1]{[#1]}%
\providecommand \BibitemOpen [0]{}%
\providecommand \bibitemStop [0]{}%
\providecommand \bibitemNoStop [0]{.\EOS\space}%
\providecommand \EOS [0]{\spacefactor3000\relax}%
\providecommand \BibitemShut  [1]{\csname bibitem#1\endcsname}%
\let\auto@bib@innerbib\@empty
\bibitem [{\citenamefont {Hohenberg}\ and\ \citenamefont {Kohn}(1964)}]{HK}%
  \BibitemOpen
  \bibfield  {author} {\bibinfo {author} {\bibfnamefont {P.}~\bibnamefont
  {Hohenberg}}\ and\ \bibinfo {author} {\bibfnamefont {W.}~\bibnamefont
  {Kohn}},\ }\href@noop {} {\bibfield  {journal} {\bibinfo  {journal} {Phys.
  Rev.}\ }\textbf {\bibinfo {volume} {136}},\ \bibinfo {pages} {B864} (\bibinfo
  {year} {1964})}\BibitemShut {NoStop}%
\bibitem [{\citenamefont {Karasiev}\ \emph
  {et~al.}(2013{\natexlab{a}})\citenamefont {Karasiev}, \citenamefont
  {D.Chakraborty},\ and\ \citenamefont {Trickey}}]{KCT2013}%
  \BibitemOpen
  \bibfield  {author} {\bibinfo {author} {\bibfnamefont {V.}~\bibnamefont
  {Karasiev}}, \bibinfo {author} {\bibnamefont {D.Chakraborty}}, \ and\
  \bibinfo {author} {\bibfnamefont {S.}~\bibnamefont {Trickey}},\ }in\
  \href@noop {} {\emph {\bibinfo {booktitle} {Many-Electron Approaches in
  Physics, Chemistry, and Mathematics}}},\ \bibinfo {editor} {edited by\
  \bibinfo {editor} {\bibfnamefont {L.~D.}\ \bibnamefont {Site}}\ and\ \bibinfo
  {editor} {\bibfnamefont {V.}~\bibnamefont {Bach}}}\ (\bibinfo  {publisher}
  {Springer Verlag},\ \bibinfo {address} {Berlin},\ \bibinfo {year}
  {2013})\BibitemShut {NoStop}%
\bibitem [{\citenamefont {Akimov}\ and\ \citenamefont
  {Prezhdo}(2015)}]{AkimovPrezhdo}%
  \BibitemOpen
  \bibfield  {author} {\bibinfo {author} {\bibfnamefont {A.~V.}\ \bibnamefont
  {Akimov}}\ and\ \bibinfo {author} {\bibfnamefont {O.~V.}\ \bibnamefont
  {Prezhdo}},\ }\href {\doibase 10.1021/cr500524c} {\bibfield  {journal}
  {\bibinfo  {journal} {Chemical Reviews}\ }\textbf {\bibinfo {volume} {115}},\
  \bibinfo {pages} {5797} (\bibinfo {year} {2015})},\ \bibinfo {note} {pMID:
  25851499}\BibitemShut {NoStop}%
\bibitem [{\citenamefont {Graziani}(2014)}]{WDM}%
  \BibitemOpen
  \bibinfo {editor} {\bibfnamefont {F.}~\bibnamefont {Graziani}},\ ed.,\
  \href@noop {} {\emph {\bibinfo {title} {Frontiers and Challenges in Warm
  Dense Matter}}}\ (\bibinfo  {publisher} {Springer Verlag},\ \bibinfo
  {address} {Berlin},\ \bibinfo {year} {2014})\BibitemShut {NoStop}%
\bibitem [{WDM()}]{WDMBasicNeeds}%
  \BibitemOpen
  \href@noop {} {\emph {\bibinfo {title} {Basic Research Needs for High Energy
  Density Laboratory Physics: Report on the Workshop on High Energy Density
  Laboratory Physics Research Needs, Nov. 15–18, 2009}}},\ \bibinfo {type}
  {Tech. Rep.}\ (\bibinfo  {institution} {Dept. of Energy})\BibitemShut
  {NoStop}%
\bibitem [{\citenamefont {Becke}(1988)}]{BeckeGGA}%
  \BibitemOpen
  \bibfield  {author} {\bibinfo {author} {\bibfnamefont {A.~D.}\ \bibnamefont
  {Becke}},\ }\href@noop {} {\bibfield  {journal} {\bibinfo  {journal} {Phys.
  Rev. A}\ }\textbf {\bibinfo {volume} {38}},\ \bibinfo {pages} {3098}
  (\bibinfo {year} {1988})}\BibitemShut {NoStop}%
\bibitem [{\citenamefont {Lee}\ \emph {et~al.}(1988)\citenamefont {Lee},
  \citenamefont {Yang},\ and\ \citenamefont {Parr}}]{LYP}%
  \BibitemOpen
  \bibfield  {author} {\bibinfo {author} {\bibfnamefont {C.}~\bibnamefont
  {Lee}}, \bibinfo {author} {\bibfnamefont {W.}~\bibnamefont {Yang}}, \ and\
  \bibinfo {author} {\bibfnamefont {R.~G.}\ \bibnamefont {Parr}},\ }\href
  {\doibase 10.1103/PhysRevB.37.785} {\bibfield  {journal} {\bibinfo  {journal}
  {Phys. Rev. B}\ }\textbf {\bibinfo {volume} {37}},\ \bibinfo {pages} {785}
  (\bibinfo {year} {1988})}\BibitemShut {NoStop}%
\bibitem [{\citenamefont {Perdew}\ \emph {et~al.}()\citenamefont {Perdew},
  \citenamefont {Burke},\ and\ \citenamefont {Ernzerhof}}]{PBE}%
  \BibitemOpen
  \bibfield  {author} {\bibinfo {author} {\bibfnamefont {J.~P.}\ \bibnamefont
  {Perdew}}, \bibinfo {author} {\bibfnamefont {K.}~\bibnamefont {Burke}}, \
  and\ \bibinfo {author} {\bibfnamefont {M.}~\bibnamefont {Ernzerhof}},\
  }\href@noop {} {}\bibinfo {note} {, Phys.\ Rev.\ Lett.\ \textbf{77}, 3865
  (1996); \textbf{78}, 1396(E) (1997).}\BibitemShut {Stop}%
\bibitem [{\citenamefont {Karasiev}\ \emph
  {et~al.}(2013{\natexlab{b}})\citenamefont {Karasiev}, \citenamefont
  {Chakraborty}, \citenamefont {Shukruto},\ and\ \citenamefont
  {Trickey}}]{KCST13}%
  \BibitemOpen
  \bibfield  {author} {\bibinfo {author} {\bibfnamefont {V.~V.}\ \bibnamefont
  {Karasiev}}, \bibinfo {author} {\bibfnamefont {D.}~\bibnamefont
  {Chakraborty}}, \bibinfo {author} {\bibfnamefont {O.~A.}\ \bibnamefont
  {Shukruto}}, \ and\ \bibinfo {author} {\bibfnamefont {S.~B.}\ \bibnamefont
  {Trickey}},\ }\href {\doibase 10.1103/PhysRevB.88.161108} {\bibfield
  {journal} {\bibinfo  {journal} {Phys. Rev. B}\ }\textbf {\bibinfo {volume}
  {88}},\ \bibinfo {pages} {161108} (\bibinfo {year}
  {2013}{\natexlab{b}})}\BibitemShut {NoStop}%
\bibitem [{\citenamefont {Karasiev}\ \emph {et~al.}(2009)\citenamefont
  {Karasiev}, \citenamefont {Jones}, \citenamefont {Trickey},\ and\
  \citenamefont {Harris}}]{KJTH09}%
  \BibitemOpen
  \bibfield  {author} {\bibinfo {author} {\bibfnamefont {V.~V.}\ \bibnamefont
  {Karasiev}}, \bibinfo {author} {\bibfnamefont {R.~S.}\ \bibnamefont {Jones}},
  \bibinfo {author} {\bibfnamefont {S.~B.}\ \bibnamefont {Trickey}}, \ and\
  \bibinfo {author} {\bibfnamefont {F.~E.}\ \bibnamefont {Harris}},\ }\href
  {\doibase 10.1103/PhysRevB.80.245120} {\bibfield  {journal} {\bibinfo
  {journal} {Phys. Rev. B}\ }\textbf {\bibinfo {volume} {80}},\ \bibinfo
  {pages} {245120} (\bibinfo {year} {2009})}\BibitemShut {NoStop}%
\bibitem [{\citenamefont {Constantin}\ \emph {et~al.}(2011)\citenamefont
  {Constantin}, \citenamefont {Fabiano}, \citenamefont {Laricchia},\ and\
  \citenamefont {Della~Sala}}]{APBE}%
  \BibitemOpen
  \bibfield  {author} {\bibinfo {author} {\bibfnamefont {L.~A.}\ \bibnamefont
  {Constantin}}, \bibinfo {author} {\bibfnamefont {E.}~\bibnamefont {Fabiano}},
  \bibinfo {author} {\bibfnamefont {S.}~\bibnamefont {Laricchia}}, \ and\
  \bibinfo {author} {\bibfnamefont {F.}~\bibnamefont {Della~Sala}},\ }\href
  {\doibase 10.1103/PhysRevLett.106.186406} {\bibfield  {journal} {\bibinfo
  {journal} {Phys. Rev. Lett.}\ }\textbf {\bibinfo {volume} {106}},\ \bibinfo
  {pages} {186406} (\bibinfo {year} {2011})}\BibitemShut {NoStop}%
\bibitem [{\citenamefont {Tran}\ and\ \citenamefont
  {Wesolowski}(2002)}]{TranWesolowski}%
  \BibitemOpen
  \bibfield  {author} {\bibinfo {author} {\bibfnamefont {F.}~\bibnamefont
  {Tran}}\ and\ \bibinfo {author} {\bibfnamefont {T.~A.}\ \bibnamefont
  {Wesolowski}},\ }\href {\doibase 10.1002/qua.10306} {\bibfield  {journal}
  {\bibinfo  {journal} {Int. J. Quantum Chem.}\ }\textbf {\bibinfo {volume}
  {89}},\ \bibinfo {pages} {441} (\bibinfo {year} {2002})}\BibitemShut
  {NoStop}%
\bibitem [{\citenamefont {Lacks}\ and\ \citenamefont
  {Gordon}(1994)}]{LacksGordon94}%
  \BibitemOpen
  \bibfield  {author} {\bibinfo {author} {\bibfnamefont {D.~J.}\ \bibnamefont
  {Lacks}}\ and\ \bibinfo {author} {\bibfnamefont {R.~G.}\ \bibnamefont
  {Gordon}},\ }\href {\doibase http://dx.doi.org/10.1063/1.466274} {\bibfield
  {journal} {\bibinfo  {journal} {J. Chem. Phys.}\ }\textbf {\bibinfo {volume}
  {100}},\ \bibinfo {pages} {4446} (\bibinfo {year} {1994})}\BibitemShut
  {NoStop}%
\bibitem [{\citenamefont {Thakkar}(1992)}]{Thakkar92}%
  \BibitemOpen
  \bibfield  {author} {\bibinfo {author} {\bibfnamefont {A.}~\bibnamefont
  {Thakkar}},\ }\href {\doibase 10.1103/PhysRevA.46.6920} {\bibfield  {journal}
  {\bibinfo  {journal} {Phys. Rev. A}\ }\textbf {\bibinfo {volume} {46}},\
  \bibinfo {pages} {6920} (\bibinfo {year} {1992})}\BibitemShut {NoStop}%
\bibitem [{\citenamefont {Thomas}(1927)}]{Thomas27}%
  \BibitemOpen
  \bibfield  {author} {\bibinfo {author} {\bibfnamefont {L.~H.}\ \bibnamefont
  {Thomas}},\ }\href {\doibase 10.1017/S0305004100011683} {\bibfield  {journal}
  {\bibinfo  {journal} {Math. Proc. Camb. Phil. Soc.}\ }\textbf {\bibinfo
  {volume} {23}},\ \bibinfo {pages} {542} (\bibinfo {year} {1927})}\BibitemShut
  {NoStop}%
\bibitem [{\citenamefont {Fermi}(1928)}]{Fermi28}%
  \BibitemOpen
  \bibfield  {author} {\bibinfo {author} {\bibfnamefont {E.}~\bibnamefont
  {Fermi}},\ }\href {http://dx.doi.org/10.1007/BF01351576} {\bibfield
  {journal} {\bibinfo  {journal} {Zeitschrift f\"ur Physik A Hadrons and
  Nuclei}\ }\textbf {\bibinfo {volume} {48}},\ \bibinfo {pages} {73} (\bibinfo
  {year} {1928})}\BibitemShut {NoStop}%
\bibitem [{\citenamefont {Karasiev}\ \emph {et~al.}(2006)\citenamefont
  {Karasiev}, \citenamefont {Trickey},\ and\ \citenamefont
  {Harris}}]{KarasievConjoint}%
  \BibitemOpen
  \bibfield  {author} {\bibinfo {author} {\bibfnamefont {V.~V.}\ \bibnamefont
  {Karasiev}}, \bibinfo {author} {\bibfnamefont {S.~B.}\ \bibnamefont
  {Trickey}}, \ and\ \bibinfo {author} {\bibfnamefont {F.~E.}\ \bibnamefont
  {Harris}},\ }\href {\doibase 10.1007/s10820-006-9019-8} {\bibfield  {journal}
  {\bibinfo  {journal} {Journal of Computer-Aided Materials Design}\ }\textbf
  {\bibinfo {volume} {13}},\ \bibinfo {pages} {111} (\bibinfo {year}
  {2006})}\BibitemShut {NoStop}%
\bibitem [{\citenamefont {Lee}\ \emph {et~al.}(1991)\citenamefont {Lee},
  \citenamefont {Lee},\ and\ \citenamefont {Parr}}]{LeeLeeParr91}%
  \BibitemOpen
  \bibfield  {author} {\bibinfo {author} {\bibfnamefont {H.}~\bibnamefont
  {Lee}}, \bibinfo {author} {\bibfnamefont {C.}~\bibnamefont {Lee}}, \ and\
  \bibinfo {author} {\bibfnamefont {R.}~\bibnamefont {Parr}},\ }\href {\doibase
  10.1103/PhysRevA.44.768} {\bibfield  {journal} {\bibinfo  {journal} {Phys.
  Rev. A}\ }\textbf {\bibinfo {volume} {44}},\ \bibinfo {pages} {768} (\bibinfo
  {year} {1991})}\BibitemShut {NoStop}%
\bibitem [{\citenamefont {Wang}\ and\ \citenamefont
  {Teter}(1992)}]{WangTeter92}%
  \BibitemOpen
  \bibfield  {author} {\bibinfo {author} {\bibfnamefont {L.-W.}\ \bibnamefont
  {Wang}}\ and\ \bibinfo {author} {\bibfnamefont {M.}~\bibnamefont {Teter}},\
  }\href {\doibase 10.1103/PhysRevB.45.13196} {\bibfield  {journal} {\bibinfo
  {journal} {Phys. Rev. B}\ }\textbf {\bibinfo {volume} {45}},\ \bibinfo
  {pages} {13196} (\bibinfo {year} {1992})}\BibitemShut {NoStop}%
\bibitem [{\citenamefont {Wang}\ \emph {et~al.}(1999)\citenamefont {Wang},
  \citenamefont {Govind},\ and\ \citenamefont {Carter}}]{WGC1999}%
  \BibitemOpen
  \bibfield  {author} {\bibinfo {author} {\bibfnamefont {Y.}~\bibnamefont
  {Wang}}, \bibinfo {author} {\bibfnamefont {N.}~\bibnamefont {Govind}}, \ and\
  \bibinfo {author} {\bibfnamefont {E.}~\bibnamefont {Carter}},\ }\href
  {\doibase 10.1103/PhysRevB.60.16350} {\bibfield  {journal} {\bibinfo
  {journal} {Phys. Rev. B}\ }\textbf {\bibinfo {volume} {60}},\ \bibinfo
  {pages} {16350} (\bibinfo {year} {1999})}\BibitemShut {NoStop}%
\bibitem [{\citenamefont {Wang}\ \emph {et~al.}(2001)\citenamefont {Wang},
  \citenamefont {Govind},\ and\ \citenamefont {Carter}}]{ErratumWGC1999}%
  \BibitemOpen
  \bibfield  {author} {\bibinfo {author} {\bibfnamefont {Y.}~\bibnamefont
  {Wang}}, \bibinfo {author} {\bibfnamefont {N.}~\bibnamefont {Govind}}, \ and\
  \bibinfo {author} {\bibfnamefont {E.}~\bibnamefont {Carter}},\ }\href
  {\doibase 10.1103/PhysRevB.64.089903} {\bibfield  {journal} {\bibinfo
  {journal} {Phys. Rev. B}\ }\textbf {\bibinfo {volume} {64}},\ \bibinfo
  {pages} {089903} (\bibinfo {year} {2001})}\BibitemShut {NoStop}%
\bibitem [{\citenamefont {Huang}\ and\ \citenamefont
  {Carter}(2010)}]{HuangCarter10}%
  \BibitemOpen
  \bibfield  {author} {\bibinfo {author} {\bibfnamefont {C.}~\bibnamefont
  {Huang}}\ and\ \bibinfo {author} {\bibfnamefont {E.}~\bibnamefont {Carter}},\
  }\href {\doibase 10.1103/PhysRevB.81.045206} {\bibfield  {journal} {\bibinfo
  {journal} {Phys. Rev. B}\ }\textbf {\bibinfo {volume} {81}},\ \bibinfo
  {pages} {045206} (\bibinfo {year} {2010})}\BibitemShut {NoStop}%
\bibitem [{\citenamefont {Ke}\ \emph {et~al.}(2014)\citenamefont {Ke},
  \citenamefont {Libisch}, \citenamefont {Xia},\ and\ \citenamefont
  {Carter}}]{KLXC14}%
  \BibitemOpen
  \bibfield  {author} {\bibinfo {author} {\bibfnamefont {Y.}~\bibnamefont
  {Ke}}, \bibinfo {author} {\bibfnamefont {F.}~\bibnamefont {Libisch}},
  \bibinfo {author} {\bibfnamefont {J.}~\bibnamefont {Xia}}, \ and\ \bibinfo
  {author} {\bibfnamefont {E.~A.}\ \bibnamefont {Carter}},\ }\href {\doibase
  10.1103/PhysRevB.89.155112} {\bibfield  {journal} {\bibinfo  {journal} {Phys.
  Rev. B}\ }\textbf {\bibinfo {volume} {89}},\ \bibinfo {pages} {155112}
  (\bibinfo {year} {2014})}\BibitemShut {NoStop}%
\bibitem [{\citenamefont {Hung}\ and\ \citenamefont
  {Carter}(2011)}]{HungCarter11}%
  \BibitemOpen
  \bibfield  {author} {\bibinfo {author} {\bibfnamefont {L.}~\bibnamefont
  {Hung}}\ and\ \bibinfo {author} {\bibfnamefont {E.~A.}\ \bibnamefont
  {Carter}},\ }\href@noop {} {\bibfield  {journal} {\bibinfo  {journal} {J.
  Phys. Chem. C}\ }\textbf {\bibinfo {volume} {115}},\ \bibinfo {pages} {6269}
  (\bibinfo {year} {2011})}\BibitemShut {NoStop}%
\bibitem [{\citenamefont {Shin}\ and\ \citenamefont
  {Carter}(2013)}]{ShinCarter13}%
  \BibitemOpen
  \bibfield  {author} {\bibinfo {author} {\bibfnamefont {I.}~\bibnamefont
  {Shin}}\ and\ \bibinfo {author} {\bibfnamefont {E.~A.}\ \bibnamefont
  {Carter}},\ }\href {\doibase 10.1103/PhysRevB.88.064106} {\bibfield
  {journal} {\bibinfo  {journal} {Phys. Rev. B}\ }\textbf {\bibinfo {volume}
  {88}},\ \bibinfo {pages} {064106} (\bibinfo {year} {2013})}\BibitemShut
  {NoStop}%
\bibitem [{\citenamefont {Lieb}\ and\ \citenamefont {Simon}(1977)}]{LiebSimon}%
  \BibitemOpen
  \bibfield  {author} {\bibinfo {author} {\bibfnamefont {E.~H.}\ \bibnamefont
  {Lieb}}\ and\ \bibinfo {author} {\bibfnamefont {B.}~\bibnamefont {Simon}},\
  }\href@noop {} {\bibfield  {journal} {\bibinfo  {journal} {Adv. in Math.}\
  }\textbf {\bibinfo {volume} {23}},\ \bibinfo {pages} {22} (\bibinfo {year}
  {1977})}\BibitemShut {NoStop}%
\bibitem [{\citenamefont {Lieb}\ and\ \citenamefont {Simon}(1973)}]{LS73}%
  \BibitemOpen
  \bibfield  {author} {\bibinfo {author} {\bibfnamefont {E.}~\bibnamefont
  {Lieb}}\ and\ \bibinfo {author} {\bibfnamefont {B.}~\bibnamefont {Simon}},\
  }\href@noop {} {\bibfield  {journal} {\bibinfo  {journal} {Phys. Rev. Lett.}\
  }\textbf {\bibinfo {volume} {31}},\ \bibinfo {pages} {681} (\bibinfo {year}
  {1973})}\BibitemShut {NoStop}%
\bibitem [{\citenamefont {Spruch}(1991)}]{Spruch}%
  \BibitemOpen
  \bibfield  {author} {\bibinfo {author} {\bibfnamefont {L.}~\bibnamefont
  {Spruch}},\ }\href@noop {} {\bibfield  {journal} {\bibinfo  {journal}
  {Reviews of Modern Physics}\ }\textbf {\bibinfo {volume} {63}} (\bibinfo
  {year} {1991})}\BibitemShut {NoStop}%
\bibitem [{\citenamefont {Scott}(1952)}]{Scott}%
  \BibitemOpen
  \bibfield  {author} {\bibinfo {author} {\bibfnamefont {J.~M.~C.}\
  \bibnamefont {Scott}},\ }\href@noop {} {\bibfield  {journal} {\bibinfo
  {journal} {Philos. Mag.}\ }\textbf {\bibinfo {volume} {43}},\ \bibinfo
  {pages} {859} (\bibinfo {year} {1952})}\BibitemShut {NoStop}%
\bibitem [{\citenamefont {March}\ and\ \citenamefont
  {Plaskett}(1956)}]{MarchPlaskett}%
  \BibitemOpen
  \bibfield  {author} {\bibinfo {author} {\bibfnamefont {N.~H.}\ \bibnamefont
  {March}}\ and\ \bibinfo {author} {\bibfnamefont {J.~S.}\ \bibnamefont
  {Plaskett}},\ }\href@noop {} {\bibfield  {journal} {\bibinfo  {journal}
  {Proceedings of the Royal Society A}\ }\textbf {\bibinfo {volume} {235}},\
  \bibinfo {pages} {419} (\bibinfo {year} {1956})}\BibitemShut {NoStop}%
\bibitem [{\citenamefont {March}\ and\ \citenamefont {Parr}(1980)}]{MarchParr}%
  \BibitemOpen
  \bibfield  {author} {\bibinfo {author} {\bibfnamefont {N.~H.}\ \bibnamefont
  {March}}\ and\ \bibinfo {author} {\bibfnamefont {R.~G.}\ \bibnamefont
  {Parr}},\ }\href@noop {} {\bibfield  {journal} {\bibinfo  {journal} {Proc.
  Natl. Acad. Sci. USA}\ }\textbf {\bibinfo {volume} {77}},\ \bibinfo {pages}
  {6285} (\bibinfo {year} {1980})}\BibitemShut {NoStop}%
\bibitem [{\citenamefont {Schwinger}(1980)}]{Schwinger80}%
  \BibitemOpen
  \bibfield  {author} {\bibinfo {author} {\bibfnamefont {J.}~\bibnamefont
  {Schwinger}},\ }\href@noop {} {\bibfield  {journal} {\bibinfo  {journal}
  {Phys. Rev. A}\ }\textbf {\bibinfo {volume} {22}},\ \bibinfo {pages} {1827}
  (\bibinfo {year} {1980})}\BibitemShut {NoStop}%
\bibitem [{\citenamefont {Englert}\ and\ \citenamefont
  {Schwinger}(1985)}]{ES85}%
  \BibitemOpen
  \bibfield  {author} {\bibinfo {author} {\bibfnamefont {B.-G.}\ \bibnamefont
  {Englert}}\ and\ \bibinfo {author} {\bibfnamefont {J.}~\bibnamefont
  {Schwinger}},\ }\href@noop {} {\bibfield  {journal} {\bibinfo  {journal}
  {Phys. Rev. A}\ }\textbf {\bibinfo {volume} {32}},\ \bibinfo {pages} {26}
  (\bibinfo {year} {1985})}\BibitemShut {NoStop}%
\bibitem [{\citenamefont {Englert}\ and\ \citenamefont
  {Schwinger}(1982)}]{ES82}%
  \BibitemOpen
  \bibfield  {author} {\bibinfo {author} {\bibfnamefont {B.-G.}\ \bibnamefont
  {Englert}}\ and\ \bibinfo {author} {\bibfnamefont {J.}~\bibnamefont
  {Schwinger}},\ }\href {\doibase 10.1103/PhysRevA.26.2322} {\bibfield
  {journal} {\bibinfo  {journal} {Phys. Rev. A}\ }\textbf {\bibinfo {volume}
  {26}},\ \bibinfo {pages} {2322} (\bibinfo {year} {1982})}\BibitemShut
  {NoStop}%
\bibitem [{\citenamefont {Elliott}\ and\ \citenamefont
  {Burke}(2009)}]{ElliottBurke}%
  \BibitemOpen
  \bibfield  {author} {\bibinfo {author} {\bibfnamefont {P.}~\bibnamefont
  {Elliott}}\ and\ \bibinfo {author} {\bibfnamefont {K.}~\bibnamefont
  {Burke}},\ }\href {\doibase 10.1139/V09-095} {\bibfield  {journal} {\bibinfo
  {journal} {Canadian Journal of Chemistry}\ }\textbf {\bibinfo {volume}
  {87}},\ \bibinfo {pages} {1485} (\bibinfo {year} {2009})}\BibitemShut
  {NoStop}%
\bibitem [{\citenamefont {Burke}\ \emph {et~al.}(2016)\citenamefont {Burke},
  \citenamefont {Cancio}, \citenamefont {Gould},\ and\ \citenamefont
  {Pittalis}}]{BCGP16}%
  \BibitemOpen
  \bibfield  {author} {\bibinfo {author} {\bibfnamefont {K.}~\bibnamefont
  {Burke}}, \bibinfo {author} {\bibfnamefont {A.}~\bibnamefont {Cancio}},
  \bibinfo {author} {\bibfnamefont {T.}~\bibnamefont {Gould}}, \ and\ \bibinfo
  {author} {\bibfnamefont {S.}~\bibnamefont {Pittalis}},\ }\href@noop {} {\
  (\bibinfo {year} {2016})},\ \bibinfo {note} {arXiv:1602.08546}\BibitemShut
  {NoStop}%
\bibitem [{\citenamefont {Burke}\ \emph {et~al.}(2014)\citenamefont {Burke},
  \citenamefont {Cancio}, \citenamefont {Gould},\ and\ \citenamefont
  {Pittalis}}]{BCGP14}%
  \BibitemOpen
  \bibfield  {author} {\bibinfo {author} {\bibfnamefont {K.}~\bibnamefont
  {Burke}}, \bibinfo {author} {\bibfnamefont {A.}~\bibnamefont {Cancio}},
  \bibinfo {author} {\bibfnamefont {T.}~\bibnamefont {Gould}}, \ and\ \bibinfo
  {author} {\bibfnamefont {S.}~\bibnamefont {Pittalis}},\ }\href@noop {} {\
  (\bibinfo {year} {2014})},\ \bibinfo {note} {arXiv:1409.4834}\BibitemShut
  {NoStop}%
\bibitem [{\citenamefont {Perdew}\ \emph {et~al.}(2008)\citenamefont {Perdew},
  \citenamefont {Ruzsinszky}, \citenamefont {Csonka}, \citenamefont {Vydrov},
  \citenamefont {Scuseria}, \citenamefont {Constantin}, \citenamefont {Zhou},\
  and\ \citenamefont {Burke}}]{PBEsol}%
  \BibitemOpen
  \bibfield  {author} {\bibinfo {author} {\bibfnamefont {J.~P.}\ \bibnamefont
  {Perdew}}, \bibinfo {author} {\bibfnamefont {A.}~\bibnamefont {Ruzsinszky}},
  \bibinfo {author} {\bibfnamefont {G.~I.}\ \bibnamefont {Csonka}}, \bibinfo
  {author} {\bibfnamefont {O.~A.}\ \bibnamefont {Vydrov}}, \bibinfo {author}
  {\bibfnamefont {G.~E.}\ \bibnamefont {Scuseria}}, \bibinfo {author}
  {\bibfnamefont {L.~A.}\ \bibnamefont {Constantin}}, \bibinfo {author}
  {\bibfnamefont {X.}~\bibnamefont {Zhou}}, \ and\ \bibinfo {author}
  {\bibfnamefont {K.}~\bibnamefont {Burke}},\ }\href {\doibase
  10.1103/PhysRevLett.100.136406} {\bibfield  {journal} {\bibinfo  {journal}
  {Phys. Rev. Lett.}\ }\textbf {\bibinfo {volume} {100}},\ \bibinfo {pages}
  {136406} (\bibinfo {year} {2008})}\BibitemShut {NoStop}%
\bibitem [{\citenamefont {Lee}\ \emph {et~al.}(2009)\citenamefont {Lee},
  \citenamefont {Constantin}, \citenamefont {Perdew},\ and\ \citenamefont
  {Burke}}]{LCPB09}%
  \BibitemOpen
  \bibfield  {author} {\bibinfo {author} {\bibfnamefont {D.}~\bibnamefont
  {Lee}}, \bibinfo {author} {\bibfnamefont {L.~A.}\ \bibnamefont {Constantin}},
  \bibinfo {author} {\bibfnamefont {J.~P.}\ \bibnamefont {Perdew}}, \ and\
  \bibinfo {author} {\bibfnamefont {K.}~\bibnamefont {Burke}},\ }\href@noop {}
  {\bibfield  {journal} {\bibinfo  {journal} {The Journal of chemical physics}\
  }\textbf {\bibinfo {volume} {130}},\ \bibinfo {pages} {034107} (\bibinfo
  {year} {2009})}\BibitemShut {NoStop}%
\bibitem [{\citenamefont {Della~Sala}\ \emph {et~al.}(2015)\citenamefont
  {Della~Sala}, \citenamefont {Fabiano},\ and\ \citenamefont
  {Constantin}}]{DSFC15}%
  \BibitemOpen
  \bibfield  {author} {\bibinfo {author} {\bibfnamefont {F.}~\bibnamefont
  {Della~Sala}}, \bibinfo {author} {\bibfnamefont {E.}~\bibnamefont {Fabiano}},
  \ and\ \bibinfo {author} {\bibfnamefont {L.~A.}\ \bibnamefont {Constantin}},\
  }\href {\doibase 10.1103/PhysRevB.91.035126} {\bibfield  {journal} {\bibinfo
  {journal} {Phys. Rev. B}\ }\textbf {\bibinfo {volume} {91}},\ \bibinfo
  {pages} {035126} (\bibinfo {year} {2015})}\BibitemShut {NoStop}%
\bibitem [{\citenamefont {Laricchia}\ \emph {et~al.}(2014)\citenamefont
  {Laricchia}, \citenamefont {Constantin}, \citenamefont {Fabiano},\ and\
  \citenamefont {Sala}}]{LCFSLapl}%
  \BibitemOpen
  \bibfield  {author} {\bibinfo {author} {\bibfnamefont {S.}~\bibnamefont
  {Laricchia}}, \bibinfo {author} {\bibfnamefont {L.~A.}\ \bibnamefont
  {Constantin}}, \bibinfo {author} {\bibfnamefont {E.}~\bibnamefont {Fabiano}},
  \ and\ \bibinfo {author} {\bibfnamefont {F.~D.}\ \bibnamefont {Sala}},\
  }\href {\doibase 10.1021/ct400836s} {\bibfield  {journal} {\bibinfo
  {journal} {Journal of Chemical Theory and Computation}\ }\textbf {\bibinfo
  {volume} {10}},\ \bibinfo {pages} {164} (\bibinfo {year} {2014})}\BibitemShut
  {NoStop}%
\bibitem [{\citenamefont {Constantin}\ \emph {et~al.}(2016)\citenamefont
  {Constantin}, \citenamefont {Fabiano},\ and\ \citenamefont
  {Della~Sala}}]{CFDS16}%
  \BibitemOpen
  \bibfield  {author} {\bibinfo {author} {\bibfnamefont {L.~A.}\ \bibnamefont
  {Constantin}}, \bibinfo {author} {\bibfnamefont {E.}~\bibnamefont {Fabiano}},
  \ and\ \bibinfo {author} {\bibfnamefont {F.}~\bibnamefont {Della~Sala}},\
  }\href {\doibase 10.3390/computation4020019} {\bibfield  {journal} {\bibinfo
  {journal} {Computation}\ }\textbf {\bibinfo {volume} {4}} (\bibinfo {year}
  {2016}),\ 10.3390/computation4020019}\BibitemShut {NoStop}%
\bibitem [{\citenamefont {Becke}\ and\ \citenamefont
  {Edgecombe}(1990)}]{BeckeELF}%
  \BibitemOpen
  \bibfield  {author} {\bibinfo {author} {\bibfnamefont {A.~D.}\ \bibnamefont
  {Becke}}\ and\ \bibinfo {author} {\bibfnamefont {K.~E.}\ \bibnamefont
  {Edgecombe}},\ }\href {http://link.aip.org/link/?JCP/92/5397/1} {\bibfield
  {journal} {\bibinfo  {journal} {J. Chem. Phys.}\ }\textbf {\bibinfo {volume}
  {92}},\ \bibinfo {pages} {5397} (\bibinfo {year} {1990})}\BibitemShut
  {NoStop}%
\bibitem [{\citenamefont {Silvi}\ and\ \citenamefont {Savin}(1994)}]{SavinELF}%
  \BibitemOpen
  \bibfield  {author} {\bibinfo {author} {\bibfnamefont {B.}~\bibnamefont
  {Silvi}}\ and\ \bibinfo {author} {\bibfnamefont {A.}~\bibnamefont {Savin}},\
  }\href@noop {} {\bibfield  {journal} {\bibinfo  {journal} {Nature (London)}\
  }\textbf {\bibinfo {volume} {371}},\ \bibinfo {pages} {683} (\bibinfo {year}
  {1994})}\BibitemShut {NoStop}%
\bibitem [{\citenamefont {Becke}(1998)}]{BeckemGGA}%
  \BibitemOpen
  \bibfield  {author} {\bibinfo {author} {\bibfnamefont {A.~D.}\ \bibnamefont
  {Becke}},\ }\href@noop {} {\bibfield  {journal} {\bibinfo  {journal} {J.
  Chem. Phys.}\ }\textbf {\bibinfo {volume} {109}},\ \bibinfo {pages} {2092}
  (\bibinfo {year} {1998})}\BibitemShut {NoStop}%
\bibitem [{\citenamefont {Sun}\ \emph {et~al.}(2013)\citenamefont {Sun},
  \citenamefont {Xiao}, \citenamefont {Fang}, \citenamefont {Haunschild},
  \citenamefont {Hao}, \citenamefont {Ruzsinszky}, \citenamefont {Csonka},
  \citenamefont {Scuseria},\ and\ \citenamefont {Perdew}}]{MGGA-MS}%
  \BibitemOpen
  \bibfield  {author} {\bibinfo {author} {\bibfnamefont {J.}~\bibnamefont
  {Sun}}, \bibinfo {author} {\bibfnamefont {B.}~\bibnamefont {Xiao}}, \bibinfo
  {author} {\bibfnamefont {Y.}~\bibnamefont {Fang}}, \bibinfo {author}
  {\bibfnamefont {R.}~\bibnamefont {Haunschild}}, \bibinfo {author}
  {\bibfnamefont {P.}~\bibnamefont {Hao}}, \bibinfo {author} {\bibfnamefont
  {A.}~\bibnamefont {Ruzsinszky}}, \bibinfo {author} {\bibfnamefont {G.~I.}\
  \bibnamefont {Csonka}}, \bibinfo {author} {\bibfnamefont {G.~E.}\
  \bibnamefont {Scuseria}}, \ and\ \bibinfo {author} {\bibfnamefont {J.~P.}\
  \bibnamefont {Perdew}},\ }\href {\doibase 10.1103/PhysRevLett.111.106401}
  {\bibfield  {journal} {\bibinfo  {journal} {Phys. Rev. Lett.}\ }\textbf
  {\bibinfo {volume} {111}},\ \bibinfo {pages} {106401} (\bibinfo {year}
  {2013})}\BibitemShut {NoStop}%
\bibitem [{\citenamefont {Cancio}\ \emph {et~al.}(2016)\citenamefont {Cancio},
  \citenamefont {Stewart},\ and\ \citenamefont {Kuna}}]{CSK16}%
  \BibitemOpen
  \bibfield  {author} {\bibinfo {author} {\bibfnamefont {A.}~\bibnamefont
  {Cancio}}, \bibinfo {author} {\bibfnamefont {D.}~\bibnamefont {Stewart}}, \
  and\ \bibinfo {author} {\bibfnamefont {A.}~\bibnamefont {Kuna}},\ }\href@noop
  {} {\bibfield  {journal} {\bibinfo  {journal} {Journal of Chemical Physics}\
  }\textbf {\bibinfo {volume} {144}},\ \bibinfo {pages} {084107} (\bibinfo
  {year} {2016})}\BibitemShut {NoStop}%
\bibitem [{\citenamefont {Finzel}(2015)}]{Finzel15}%
  \BibitemOpen
  \bibfield  {author} {\bibinfo {author} {\bibfnamefont {K.}~\bibnamefont
  {Finzel}},\ }\href@noop {} {\bibfield  {journal} {\bibinfo  {journal}
  {Theoretical Chemistry Accounts}\ }\textbf {\bibinfo {volume} {134}},\
  \bibinfo {eid} {106} (\bibinfo {year} {2015})}\BibitemShut {NoStop}%
\bibitem [{\citenamefont {Xia}\ and\ \citenamefont
  {Carter}(2015{\natexlab{a}})}]{XC15}%
  \BibitemOpen
  \bibfield  {author} {\bibinfo {author} {\bibfnamefont {J.}~\bibnamefont
  {Xia}}\ and\ \bibinfo {author} {\bibfnamefont {E.~A.}\ \bibnamefont
  {Carter}},\ }\href {\doibase 10.1103/PhysRevB.91.045124} {\bibfield
  {journal} {\bibinfo  {journal} {Phys. Rev. B}\ }\textbf {\bibinfo {volume}
  {91}},\ \bibinfo {pages} {045124} (\bibinfo {year}
  {2015}{\natexlab{a}})}\BibitemShut {NoStop}%
\bibitem [{\citenamefont {Trickey}\ \emph {et~al.}(2015)\citenamefont
  {Trickey}, \citenamefont {Karasiev},\ and\ \citenamefont
  {Chakraborty}}]{XC15comment}%
  \BibitemOpen
  \bibfield  {author} {\bibinfo {author} {\bibfnamefont {S.~B.}\ \bibnamefont
  {Trickey}}, \bibinfo {author} {\bibfnamefont {V.~V.}\ \bibnamefont
  {Karasiev}}, \ and\ \bibinfo {author} {\bibfnamefont {D.}~\bibnamefont
  {Chakraborty}},\ }\href {\doibase 10.1103/PhysRevB.92.117101} {\bibfield
  {journal} {\bibinfo  {journal} {Phys. Rev. B}\ }\textbf {\bibinfo {volume}
  {92}},\ \bibinfo {pages} {117101} (\bibinfo {year} {2015})}\BibitemShut
  {NoStop}%
\bibitem [{\citenamefont {Xia}\ and\ \citenamefont
  {Carter}(2015{\natexlab{b}})}]{XC15reply}%
  \BibitemOpen
  \bibfield  {author} {\bibinfo {author} {\bibfnamefont {J.}~\bibnamefont
  {Xia}}\ and\ \bibinfo {author} {\bibfnamefont {E.~A.}\ \bibnamefont
  {Carter}},\ }\href {\doibase 10.1103/PhysRevB.92.117102} {\bibfield
  {journal} {\bibinfo  {journal} {Phys. Rev. B}\ }\textbf {\bibinfo {volume}
  {92}},\ \bibinfo {pages} {117102} (\bibinfo {year}
  {2015}{\natexlab{b}})}\BibitemShut {NoStop}%
\bibitem [{\citenamefont {Lindmaa}\ \emph {et~al.}(2014)\citenamefont
  {Lindmaa}, \citenamefont {Mattsson},\ and\ \citenamefont {Armiento}}]{LAM14}%
  \BibitemOpen
  \bibfield  {author} {\bibinfo {author} {\bibfnamefont {A.}~\bibnamefont
  {Lindmaa}}, \bibinfo {author} {\bibfnamefont {A.~E.}\ \bibnamefont
  {Mattsson}}, \ and\ \bibinfo {author} {\bibfnamefont {R.}~\bibnamefont
  {Armiento}},\ }\href {\doibase 10.1103/PhysRevB.90.075139} {\bibfield
  {journal} {\bibinfo  {journal} {Phys. Rev. B}\ }\textbf {\bibinfo {volume}
  {90}},\ \bibinfo {pages} {075139} (\bibinfo {year} {2014})}\BibitemShut
  {NoStop}%
\bibitem [{\citenamefont {Perdew}\ and\ \citenamefont
  {Constantin}(2007)}]{PerdewConstantin}%
  \BibitemOpen
  \bibfield  {author} {\bibinfo {author} {\bibfnamefont {J.~P.}\ \bibnamefont
  {Perdew}}\ and\ \bibinfo {author} {\bibfnamefont {L.~A.}\ \bibnamefont
  {Constantin}},\ }\href@noop {} {\bibfield  {journal} {\bibinfo  {journal}
  {Phys. Rev. B}\ }\textbf {\bibinfo {volume} {75}},\ \bibinfo {pages}
  {155109/1} (\bibinfo {year} {2007})}\BibitemShut {NoStop}%
\bibitem [{\citenamefont {Acharya}\ \emph {et~al.}(1980)\citenamefont
  {Acharya}, \citenamefont {Bartolotti}, \citenamefont {Sears}, ,\ and\
  \citenamefont {Parr}}]{Acharya}%
  \BibitemOpen
  \bibfield  {author} {\bibinfo {author} {\bibfnamefont {P.~K.}\ \bibnamefont
  {Acharya}}, \bibinfo {author} {\bibfnamefont {L.~J.}\ \bibnamefont
  {Bartolotti}}, \bibinfo {author} {\bibfnamefont {S.~B.}\ \bibnamefont
  {Sears}}, , \ and\ \bibinfo {author} {\bibfnamefont {R.~G.}\ \bibnamefont
  {Parr}},\ }\href@noop {} {\bibfield  {journal} {\bibinfo  {journal} {Proc.
  Nati. Acad. Sci. USA}\ }\textbf {\bibinfo {volume} {77}},\ \bibinfo {pages}
  {6978} (\bibinfo {year} {1980})}\BibitemShut {NoStop}%
\bibitem [{\citenamefont {Yang}\ \emph {et~al.}(1986)\citenamefont {Yang},
  \citenamefont {Parr},\ and\ \citenamefont {Lee}}]{YPL86}%
  \BibitemOpen
  \bibfield  {author} {\bibinfo {author} {\bibfnamefont {W.}~\bibnamefont
  {Yang}}, \bibinfo {author} {\bibfnamefont {R.~G.}\ \bibnamefont {Parr}}, \
  and\ \bibinfo {author} {\bibfnamefont {C.}~\bibnamefont {Lee}},\ }\href
  {\doibase 10.1103/PhysRevA.34.4586} {\bibfield  {journal} {\bibinfo
  {journal} {Phys. Rev. A}\ }\textbf {\bibinfo {volume} {34}},\ \bibinfo
  {pages} {4586} (\bibinfo {year} {1986})}\BibitemShut {NoStop}%
\bibitem [{\citenamefont {Jones}\ and\ \citenamefont {Gunnarsson}(1989)}]{JG}%
  \BibitemOpen
  \bibfield  {author} {\bibinfo {author} {\bibfnamefont {R.~O.}\ \bibnamefont
  {Jones}}\ and\ \bibinfo {author} {\bibfnamefont {O.}~\bibnamefont
  {Gunnarsson}},\ }\href@noop {} {\bibfield  {journal} {\bibinfo  {journal}
  {Rev. Mod. Phys.}\ }\textbf {\bibinfo {volume} {61}},\ \bibinfo {pages} {689}
  (\bibinfo {year} {1989})}\BibitemShut {NoStop}%
\bibitem [{\citenamefont {Kirzhnits}(1957)}]{Kirzhnits57}%
  \BibitemOpen
  \bibfield  {author} {\bibinfo {author} {\bibfnamefont {D.}~\bibnamefont
  {Kirzhnits}},\ }\href@noop {} {\bibfield  {journal} {\bibinfo  {journal}
  {Sov. Phys. JETP}\ }\textbf {\bibinfo {volume} {5}},\ \bibinfo {pages} {64}
  (\bibinfo {year} {1957})}\BibitemShut {NoStop}%
\bibitem [{\citenamefont {Brack}\ \emph {et~al.}(1976)\citenamefont {Brack},
  \citenamefont {Jennings},\ and\ \citenamefont {Chu}}]{BrackJenningsChu}%
  \BibitemOpen
  \bibfield  {author} {\bibinfo {author} {\bibfnamefont {M.}~\bibnamefont
  {Brack}}, \bibinfo {author} {\bibfnamefont {B.~K.}\ \bibnamefont {Jennings}},
  \ and\ \bibinfo {author} {\bibfnamefont {Y.~H.}\ \bibnamefont {Chu}},\
  }\href@noop {} {\bibfield  {journal} {\bibinfo  {journal} {Phys.\ Lett.\
  \textbf{65B}}\ }\textbf {\bibinfo {volume} {65}},\ \bibinfo {pages} {1}
  (\bibinfo {year} {1976})}\BibitemShut {NoStop}%
\bibitem [{\citenamefont {Hodges}(1973)}]{Hodges}%
  \BibitemOpen
  \bibfield  {author} {\bibinfo {author} {\bibfnamefont {C.~H.}\ \bibnamefont
  {Hodges}},\ }\href@noop {} {\bibfield  {journal} {\bibinfo  {journal} {Can.
  J. Phys.}\ }\textbf {\bibinfo {volume} {51}},\ \bibinfo {pages} {1428}
  (\bibinfo {year} {1973})}\BibitemShut {NoStop}%
\bibitem [{\citenamefont {Murphy}(1981)}]{Murphy81}%
  \BibitemOpen
  \bibfield  {author} {\bibinfo {author} {\bibfnamefont {D.}~\bibnamefont
  {Murphy}},\ }\href@noop {} {\bibfield  {journal} {\bibinfo  {journal} {Phys.
  Rev. A}\ }\textbf {\bibinfo {volume} {24}},\ \bibinfo {pages} {1682}
  (\bibinfo {year} {1981})}\BibitemShut {NoStop}%
\bibitem [{\citenamefont {de~Silva}\ and\ \citenamefont
  {Corminboeuf}(2015)}]{dSC15}%
  \BibitemOpen
  \bibfield  {author} {\bibinfo {author} {\bibfnamefont {P.}~\bibnamefont
  {de~Silva}}\ and\ \bibinfo {author} {\bibfnamefont {C.}~\bibnamefont
  {Corminboeuf}},\ }\href {\doibase http://dx.doi.org/10.1063/1.4931628}
  {\bibfield  {journal} {\bibinfo  {journal} {J. Chem. Phys.}\ }\textbf
  {\bibinfo {volume} {143}} (\bibinfo {year} {2015}),\
  http://dx.doi.org/10.1063/1.4931628}\BibitemShut {NoStop}%
\bibitem [{\citenamefont {Weizs\"acker}(1935)}]{vonWeizsacker}%
  \BibitemOpen
  \bibfield  {author} {\bibinfo {author} {\bibfnamefont {C.}~\bibnamefont
  {Weizs\"acker}},\ }\href {\doibase 10.1007/BF01337700} {\bibfield  {journal}
  {\bibinfo  {journal} {Zeitschrift f\"ur Physik}\ }\textbf {\bibinfo {volume}
  {96}},\ \bibinfo {pages} {431} (\bibinfo {year} {1935})}\BibitemShut
  {NoStop}%
\bibitem [{\citenamefont {Herring}(1986)}]{Herring86}%
  \BibitemOpen
  \bibfield  {author} {\bibinfo {author} {\bibfnamefont {C.}~\bibnamefont
  {Herring}},\ }\href {\doibase 10.1103/PhysRevA.34.2614} {\bibfield  {journal}
  {\bibinfo  {journal} {Phys. Rev. A}\ }\textbf {\bibinfo {volume} {34}},\
  \bibinfo {pages} {2614} (\bibinfo {year} {1986})}\BibitemShut {NoStop}%
\bibitem [{\citenamefont {Levy}\ and\ \citenamefont {Ou-Yang}(1988)}]{LevyHui}%
  \BibitemOpen
  \bibfield  {author} {\bibinfo {author} {\bibfnamefont {M.}~\bibnamefont
  {Levy}}\ and\ \bibinfo {author} {\bibfnamefont {H.}~\bibnamefont {Ou-Yang}},\
  }\href {\doibase 10.1103/PhysRevA.38.625} {\bibfield  {journal} {\bibinfo
  {journal} {Phys. Rev. A}\ }\textbf {\bibinfo {volume} {38}},\ \bibinfo
  {pages} {625} (\bibinfo {year} {1988})}\BibitemShut {NoStop}%
\bibitem [{\citenamefont {Hao}\ \emph {et~al.}(2014)\citenamefont {Hao},
  \citenamefont {Armiento},\ and\ \citenamefont {Mattsson}}]{Fam14}%
  \BibitemOpen
  \bibfield  {author} {\bibinfo {author} {\bibfnamefont {F.}~\bibnamefont
  {Hao}}, \bibinfo {author} {\bibfnamefont {R.}~\bibnamefont {Armiento}}, \
  and\ \bibinfo {author} {\bibfnamefont {A.~E.}\ \bibnamefont {Mattsson}},\
  }\href {\doibase http://dx.doi.org/10.1063/1.4871738} {\bibfield  {journal}
  {\bibinfo  {journal} {J. Chem. Phys.}\ }\textbf {\bibinfo {volume} {140}},\
  \bibinfo {pages} {18A536} (\bibinfo {year} {2014})}\BibitemShut {NoStop}%
\bibitem [{\citenamefont {Cancio}\ and\ \citenamefont
  {Wagner}(2013)}]{CancioExqlaplFull}%
  \BibitemOpen
  \bibfield  {author} {\bibinfo {author} {\bibfnamefont {A.~C.}\ \bibnamefont
  {Cancio}}\ and\ \bibinfo {author} {\bibfnamefont {C.~E.}\ \bibnamefont
  {Wagner}},\ }\href@noop {} {\  (\bibinfo {year} {2013})},\ \bibinfo {note}
  {arXiv:1308.3744 [physics.chem-ph]}\BibitemShut {NoStop}%
\bibitem [{\citenamefont {Kato}(1957)}]{Kato57}%
  \BibitemOpen
  \bibfield  {author} {\bibinfo {author} {\bibfnamefont {T.}~\bibnamefont
  {Kato}},\ }\href@noop {} {\bibfield  {journal} {\bibinfo  {journal} {Commun.
  Pure Appl. Math.}\ }\textbf {\bibinfo {volume} {10}},\ \bibinfo {pages} {151}
  (\bibinfo {year} {1957})}\BibitemShut {NoStop}%
\bibitem [{\citenamefont {Fuchs}\ and\ \citenamefont
  {Scheffler}(1999)}]{FHI98PP}%
  \BibitemOpen
  \bibfield  {author} {\bibinfo {author} {\bibfnamefont {M.}~\bibnamefont
  {Fuchs}}\ and\ \bibinfo {author} {\bibfnamefont {M.}~\bibnamefont
  {Scheffler}},\ }\href@noop {} {\bibfield  {journal} {\bibinfo  {journal}
  {Computer Physics Communications}\ }\textbf {\bibinfo {volume} {119}},\
  \bibinfo {pages} {67} (\bibinfo {year} {1999})}\BibitemShut {NoStop}%
\bibitem [{\citenamefont {Redd}(2015)}]{Redd}%
  \BibitemOpen
  \bibfield  {author} {\bibinfo {author} {\bibfnamefont {J.~J.}\ \bibnamefont
  {Redd}},\ }\href@noop {} {Master's thesis} (\bibinfo {year}
  {2015})\BibitemShut {NoStop}%
\bibitem [{\citenamefont {Bader}(1991)}]{BaderReview}%
  \BibitemOpen
  \bibfield  {author} {\bibinfo {author} {\bibfnamefont {R.~F.~W.}\
  \bibnamefont {Bader}},\ }\href {http://dx.doi.org/10.1021/cr00005a013}
  {\bibfield  {journal} {\bibinfo  {journal} {Chemical Reviews}\ }\textbf
  {\bibinfo {volume} {91}},\ \bibinfo {pages} {893} (\bibinfo {year}
  {1991})}\BibitemShut {NoStop}%
\bibitem [{\citenamefont {Bader}\ and\ \citenamefont
  {Ess\'en}(1984)}]{BaderEssen84}%
  \BibitemOpen
  \bibfield  {author} {\bibinfo {author} {\bibfnamefont {R.~F.~W.}\
  \bibnamefont {Bader}}\ and\ \bibinfo {author} {\bibfnamefont
  {H.}~\bibnamefont {Ess\'en}},\ }\href {\doibase
  http://dx.doi.org/10.1063/1.446956} {\bibfield  {journal} {\bibinfo
  {journal} {J.\ Chem.\ Phys.}\ }\textbf {\bibinfo {volume} {80}},\ \bibinfo
  {pages} {1943} (\bibinfo {year} {1984})}\BibitemShut {NoStop}%
\end{thebibliography}

\end{document}